\documentclass[prd,preprint,superscriptaddress]{revtex4}
\usepackage{amsmath}
\usepackage{amssymb}
\usepackage{graphicx}

\newcommand{\be}{\begin{equation}}
\newcommand{\ee}{\end{equation}}
\newcommand{\ben}{\begin{eqnarray}}
\newcommand{\een}{\end{eqnarray}}

\begin{document}
\title{Limits on the parameters of the equation of state for interacting dark energy}
\author{Germ\'{a}n Izquierdo \thanks{%
E-mail address: german.izquierdo@gmail.com}}
\affiliation{Institut de Math\'ematiques de Bourgogne, Universit\'e de Bourgogne,
9 Av. Alain Savary, 21078 Dijon Cedex, France} \author{Diego Pav\'{o}n\thanks{%
E-mail address: diego.pavon@uab.es}} \affiliation{Departamento de
F\'{\i}sica, Universidad Aut\'{o}noma de Barcelona, 08193
Bellaterra (Barcelona), Spain}

\begin{abstract}
Under the assumption that cold dark matter and dark energy
interact with each other through a small coupling term, $Q$, we
constrain the parameter space of the equation of state $w$ of
those dark energy fields whose variation of the field since last
scattering do  not exceed Planck's mass. We use three
parameterizations of $w$ and two different expressions for $Q$.
Our work extends previous ones.
\end{abstract}

 \maketitle

\section{Introduction}
The equation of state parameter of dark energy -the mysterious
agent responsible for the current accelerated expansion of the
Universe- ranks among the biggest unknowns in cosmology. It is
usually written as $w= p_{\phi}/\rho_{\phi}$ and lies in the range
$(-1, -1/3)$ if the dark energy is quintessence, and $(- \infty,
-1)$ if it is of phantom type; in either case it may run with
expansion. By contrast, in the very particular (but
observationally favored) instance of a cosmological constant $\,
w$ stays fixed at $\, -1$. It is then apparent that some knowledge
on the nature of dark energy can be gained by setting limits on
$w$.

Observational constraints on this quantity from supernovae type Ia
(SN Ia), temperature anisotropies of the cosmic microwave
background radiation (CMB), baryon acoustic oscillations (BAO),
etc, suggest that nowadays $w$ is not far from the cosmological
constant value. However, all these studies rest on some or other
world model and on inevitable priors imposed upon the parameters
space to ease the data analysis.

Recently, bounds on $w$ have been derived  from the very
reasonable requirement that the variation experienced by the dark
energy scalar field $\phi$ (regardless it may be phantom or
quintessence) from any redshift, $z$, within the classical
expansion era, till now should not exceed Planck's mass (see Huang
\cite{huang}, and Saridakis \cite{saridakis}). This is to say
\begin{equation}
|\Delta \phi(z)|/M_{P} < 1 \, ,
\label{condition}
\end{equation}
\noindent where -as can be readily deduced-
\begin{equation}
\frac{|\Delta \phi (z)|}{M_{P}} = \frac{1}{M_{P}} \,
\int_{t}^{t_{0}} {\dot{\phi} \, dt} = \int_{0}^{z}{\frac{\sqrt{3
\, |1+w(x)| \, \Omega_{\phi}(x)}}{1+x}}\, dx \, . \label{Deltaphi}
\end{equation}

Huang's analysis confines itself to quintessence fields with
various $w$ parameterizations (viz: $w$ constant, a linear
function of redshift, and the Chevallier-Polarski-Linder
parametrization \cite{ch-p,linder}) \cite{huang}. Saridakis
considers phantom fields only and uses identical parameterizations
than Huang plus a further one in which $w$ depends linearly  on
the logarithm of the scale factor \cite{saridakis}.

The bound (\ref{condition}) looks a rather natural condition and
persuasive arguments  have been advanced in its favor
\cite{banks,huang1,yun-gui}. Moreover, we have found a further,
solid, motivation for it. Namely: if condition (\ref{condition})
is violated, then dark energy dominates at very early times which
cannot be reconciled with observation (see Fig. \ref{fig4} and
``forbidden" regions on the top right corner of the panels of
figures \ref{fig6}, \ref{fig9}, and \ref{fig11}, below). In
particular, as noted by Bean {\it et al.} \cite{rachelbean}, the
fractional density of dark energy cannot exceed $5 \%$ and $39 \%$
at the primeval nucleosynthesis and recombination epochs,
respectively, -we shall elaborate on this in a future publication.
Nevertheless, we refrain ourselves from imposing (\ref{condition})
on every dark energy field. Its seems safer to demand that
$|\Delta \phi|/M_{P}$  be not much larger than order unity.
However, the latter requirement cannot be easily implemented and,
at any rate, it would translate on rather loose constraints on
$w$. Therefore, we choose to circumscribe our analysis to dark
energy fields that fulfill condition (\ref{condition}).

In Huang's as well as in Saridakis' work matter and dark energy
interact with each other only gravitationally. Certainly this is a
reasonable assumption but nonetheless minimalist since, while dark
energy interactions with baryons are severely restricted by solar
system experiments \cite{solar}, there is nothing against a
possible coupling (interaction) with dark matter. On the contrary,
a transfer from dark energy to dark matter alleviates the cosmic
coincidence problem (the observational fact that both energy
densities are comparable today \cite{paul}) and may  solve it in
full \cite{rapidc}. Besides, as revealed by optical, x-ray and
weak-lensing data, the internal dynamics of 33 relaxed galaxy
clusters seems to favor such interaction \cite{binw}. Likewise, a
recent analysis using the 397 SN Ia of the ``Constitution" set
\cite{constitution}, BAO and CMB data adds significant weight to
the existence of the coupling \cite{shafieloo}. In fact, the
subject has evolved into a field of active research -see e.g.
\cite{active,gabriela1} and references therein; a recent review
can be found in \cite{review}.

In the presence of a coupling, say $Q$, the continuity equations
for the three main contributors to the present cosmic budget adopt
the form,
\begin{eqnarray}
\dot{\rho}_{b}\, &+&\, 3H \rho_{b} = 0 \, , \nonumber \\
\dot{\rho}_{c}\, &+& \, 3 H \rho_{c} = Q \, , \nonumber \\
\dot{\rho}_{\phi}\, &+& \, 3 H (1+w) \rho_{\phi} = - Q \, ,
\label{continuity}
\end{eqnarray}
where the subscripts, $b$, $c$, and $\phi$ stands for baryons,
cold dark matter, and dark energy, respectively. This paper aims
to constrain  different parameterizations of $w$ under diverse
expressions for $Q$. Regrettably, guidelines about the latter are
rather loose: it must be positive-semidefinite and small. If $Q$
were negative, the energy densities could become negative, and the
second law of thermodynamics get violated \cite{db}. If it were
large, dark energy would have dominated the expansion at early
times and possibly not today. Thus, for the time being, we must
content ourselves with guessing plausible expressions for $Q$ just
on phenomenological bases.

In view of the above equations the coupling must be a function of
$H \rho_{c}$ and $H \rho_{\phi}$. By power law expanding it up to
first order on these quantities we write $Q(H \rho_{c}, H
\rho_{\phi}) \simeq \epsilon_{c}\, H \rho_{c} \, + \,
\epsilon_{\phi}\, H \rho_{\phi}$, where both $\epsilon$
coefficients ought to be non-negative and small (i.e., not larger
than, say, $10^{-1}$). To simplify the analysis we will consider
in turn that one of the two coefficients vanishes. So, we will
take \be Q = 3 \, \epsilon H \rho_{\phi}\, , \qquad {\rm and}
\qquad Q = 3 \, \epsilon H \rho_{c} \, , \label{int} \ee -the
factor $3$ being introduced for mathematical convenience.

At this point one may object that if $Q$ is small, no significant
difference with the results of Huang \cite{huang} and Saridakis
\cite{saridakis} should  be expected. However, as we shall see,
this is not the case. {\em A priori} we can argue in favor of some
non small departure from the findings of \cite{huang} and
\cite{saridakis} by considering the ratio between $Q$ and the
second term in last equation of (\ref{continuity}). Indeed,
dividing the right hand side of (\ref{int}.1) by the second term
of the said equation yields $\epsilon/(1+w)$. In general, this
ratio cannot be neglected and it can be large in absolute value
when $w$ is close to $-1$. Likewise, using  instead (\ref{int}.2)
we have $\epsilon \, \rho_{c}/(1+w)\rho_{\phi}$. Its absolute
value can be of order unity, or even larger, for extended periods
of the cosmic history.

Although  expressions (\ref{int}) were proposed on
phenomenological grounds they can be obtained in scalar-tensor
gravity from the action (in the Einstein frame) \cite{kaloper}
\begin{equation}
S = \int{d^{4}x\, \sqrt{-g}}\, \left\{ \frac{1}{16 \pi G} R -
\frac{1}{2} \,\partial_{a}\phi \,\partial^{a} \phi \, + \,
\frac{1}{\chi^{2}(\phi)}\, L_{c}(\zeta, \partial \zeta, \chi^{-1}
g_{ab})\, + \, L_{b}(\xi, \partial \xi, \chi^{-1} g_{ab}) \right\}
\, . \label{action}
\end{equation}
Here $R$ denotes the Ricci scalar, $L_{c}$ and $\zeta$ stand for
the dark matter Lagrangian and the dark matter degrees of freedom,
respectively (corresponding meanings have $L_{b}$ and $\xi$); on
the other hand, $\chi(\phi)$ couples dark matter with the dark
energy field $\phi$. From (\ref{action}) the interaction term
between dark matter and dark energy can be expressed as
\begin{equation}
Q = H \rho_{c} \left[\frac{d(\ln \bar{\chi}(a))}{d \ln a}\right]
\, ,
\label{Q-action}
\end{equation}
where $\bar{\chi}(a) = \chi(a)^{(3w_{c} - 1)/2}$ with $w_{c}$ the
equation of state parameter of dark matter (zero in our case). By
choosing
\begin{equation}
\bar{\chi}(a) = \bar{\chi}_{0} \, \exp \left[ 3 \, \int
\frac{\epsilon_{\zeta}\,  \rho_{\zeta}\, + \,\epsilon_{c} \,
\rho_{c}} {\rho_{c}} \; d \ln a \right] \, , \label{bychoosing}
\end{equation}
the first (second) expressions for $Q$ in (\ref{int}) follows
after using (\ref{Q-action}) and setting $\epsilon_{c} = 0$
($\epsilon_{\zeta} = 0$). Similar expressions for $Q$ have been
obtained from the above action in Refs. \cite{curbelo} and
\cite{h-zhang}.

In this paper we shall assume a spatially flat
Friedmann-Robertson-Walker universe with present fractional
densities of baryons and cold dark mater $\Omega_{b0} = 0.04$, and
$\Omega_{c0} = 0.24$, respectively, throughout. As usual, a zero
subindex means the current value of the corresponding quantity;
likewise we normalize the scale factor of the metric by setting
$a_{0} = 1$.

\section{Interaction term proportional to the dark energy density}
In this section  we consider $Q = 3 \, \epsilon H \rho_{\phi}$
alongside different expressions for the equation of state
parameter.

\subsection{Constant $w$}
\begin{figure}[tbp]
\includegraphics*[scale=0.3,angle=0]{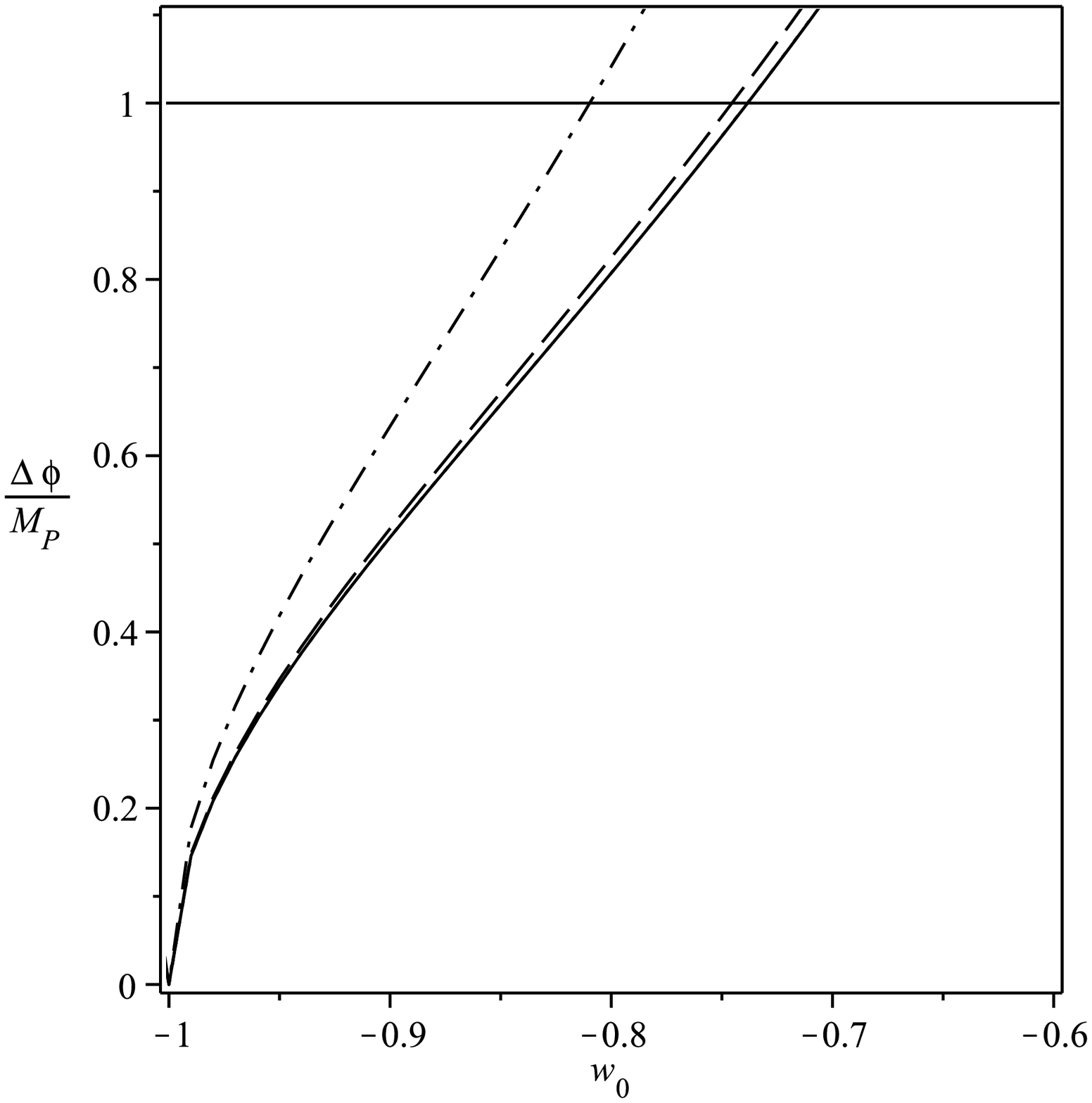}
\includegraphics*[scale=0.3,angle=0]{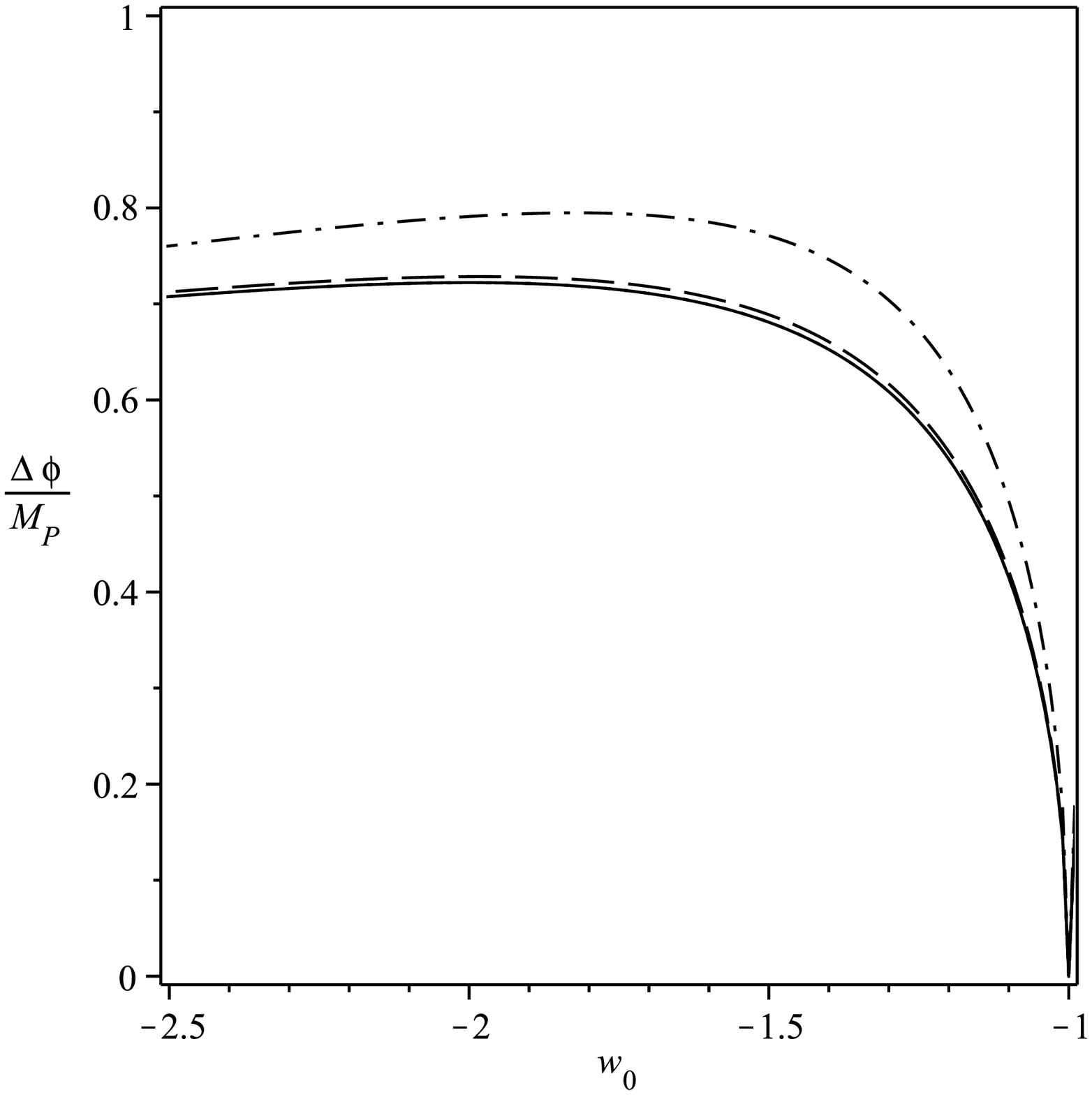}
\caption{ $|\Delta \phi|/ M_P$ vs. $w_0$  for $Q$ given by
(\ref{int}.1) and constant $w$ evaluated in the redshift range
$(0, 1089)$. Left Panel: Quintessence solutions ($-1<w_0<-1/3$).
Right panel: Phantom solutions ($w_0<-1$). In both panels solid,
dotted, dashed, and dot-dashed lines correspond to $\epsilon=0,
10^{-4}, 10^{-2},$ and $0.1$, respectively. (Solid and dotted
lines practically overlap). Condition (\ref{condition}) for
$\epsilon = 0, 10^{-4}, 10^{-2},$ and $0.1$ implies $w_{0} <
-0.738, -0.738, -0.745,$ and  $-0.810$, respectively. In drawing
these and all subsequent figures we assumed $\Omega_{b0} = 0.04$,
and $\Omega_{c0} = 0.24$.} \label{fig1}
\end{figure}

By plugging $Q=3H \epsilon \rho_{\phi}$ into (\ref{continuity}),
assuming $w = w_{0}$, and integrating we get the energy densities
dependence on the scale factor
\ben
\rho_{b}&=& \rho_{b0} \, a^{-3} \, , \nonumber \\
\rho_{c}&=& \rho_{c0}\, a^{-3}+\frac{\epsilon}{w_0+\epsilon}\,
\rho_{\phi0} \,
a^{-3}\left[1-a^{-3(w_0+\epsilon)}\right] \, , \nonumber \\
\rho_{\phi}&=&\rho_{\phi0} \, a^{-3(1+w_0+\epsilon)} \, .
\label{rho(a)1} \een We then introduce $\Omega_{\phi} =
\rho_{\phi}/(\rho_{b}\, +\, \rho_{c}\,+\, \rho_{\phi})$ in
(\ref{Deltaphi}) and numerically determine $|\Delta \phi|/ M_P$ in
the redshift interval between the last scattering $(z_{ls}= 1089)$
and now $(z = 0)$ -bear in mind that $1+z = a^{-1}$. Figure
\ref{fig1} shows $|\Delta \phi|/ M_P$ vs. $w_0$ for different
values of the parameter $\epsilon$. When $-1< w_0 < -1/3$ (left
panel), for every $\epsilon$ value there is a range of $w_{0}$
that violates condition (\ref{condition}); the bigger the
interaction strength, the bigger the $|w_{0}|$  that violates the
condition. By contrast, when $w_{0}<-1$ (right panel) the bound
(\ref{condition}) is fulfilled no matter the value of $\epsilon$.
This could have been anticipated since, in this simple case, Eq.
(\ref{continuity}.2) can be written as $\dot{\rho}_{\phi} +
3H(1+w_{eff})\, \rho_{\phi} = 0$, i.e., as though it were no
interaction but with $w$ replaced by the effective equation of
state parameter $w_{eff} = w + \epsilon$. At any rate, our results
corroborate and extend Huang's and Saridakis'.

\subsection{Chevallier-Polarski-Linder parametrization}
Except when dark energy is given by the  quantum vacuum, there is
no compelling motivation to assume $w$ constant for the whole
cosmic evolution. However, its simplest generalization in terms of
redshift, $w(z) = w_{0} + w_{1} \, z$, is not compatible with
observation for it diverges as $z \rightarrow \infty$. This
prompted the introduction of the more suitable expression $w(z)=
w_{0}\, +\, w_{1}\, \frac{z}{1+z}$ or, equivalently, in terms of
the scale factor \be w(a)= w_{0}\, +\, w_{1}(1-a)
\label{chevallier} \ee by Chevallier and Polarski \cite{ch-p}
(later popularized by Linder \cite{linder}) which does not suffer
from that drawback and behaves nearly linear in $z$ at low
redshifts. (Clearly, $w_{0} = w(z =0)$ and $w_{1} = dw(z)/dz|_{z =
0}$).

\begin{figure}[tbp]
\includegraphics*[scale=0.4,angle=0]{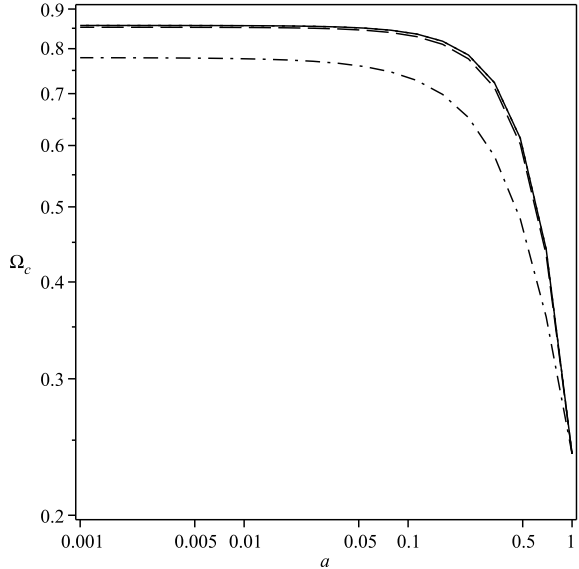}
\caption{Evolution of the fractional density of cold dark matter
$\Omega_{c} = \rho_{c}/(\rho_{b}+\rho_{c}+\rho_{\phi})$, since
last scattering till now, in terms of the scale factor for the
Chevallier-Polarski-Linder parametrization (\ref{chevallier}) with
$w_{0} = -1$ and $w_{1} = 0.5$. Solid, dotted, dashed, and
dot-dashed lines correspond to $\epsilon=0,10^{-4}, 10^{-2},$ and
$0.1$, respectively. The two first lines are undistinguishable
from one another.} \label{fig2}
\end{figure}

Now the  dark energy density integrates to
$\rho_{\phi}=\rho_{\phi0} \,a^{-3(1+w_0+w_1+\epsilon) }\,
\exp{[3w_{1} \,(a-1)]}$. By contrast, the energy density of cold
dark matter has no analytical expression but it can be found by
numerical integration of Eq. (\ref{continuity}.2). For the sake of
illustration we present in Fig. \ref{fig2}  the evolution of its
fractional density for a certain choice of parameters.
Obviously, the expression for the baryon density does not vary.

\begin{figure}[tbp]
\includegraphics*[scale=0.3,angle=0]{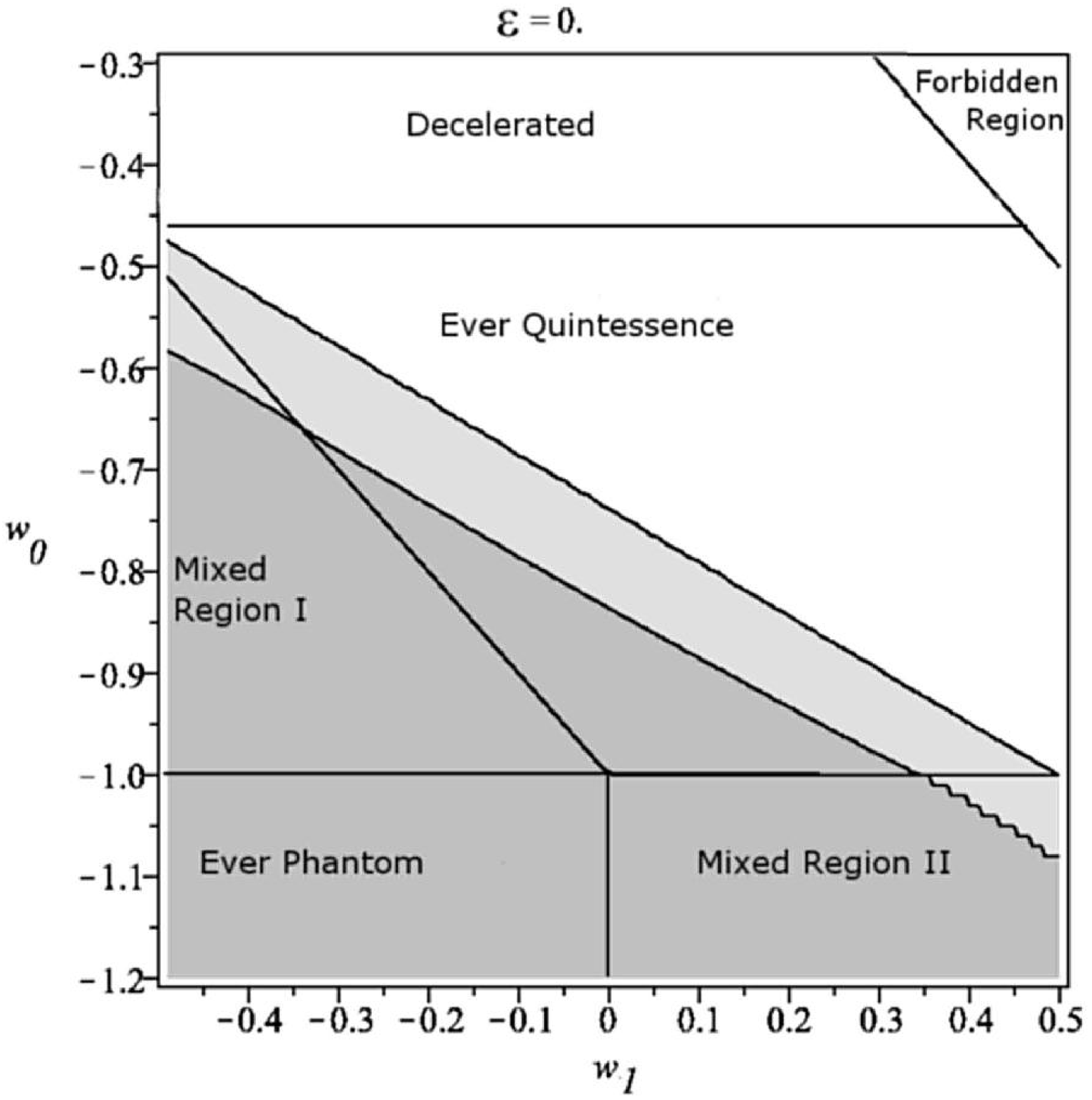}
\includegraphics*[scale=0.3,angle=0]{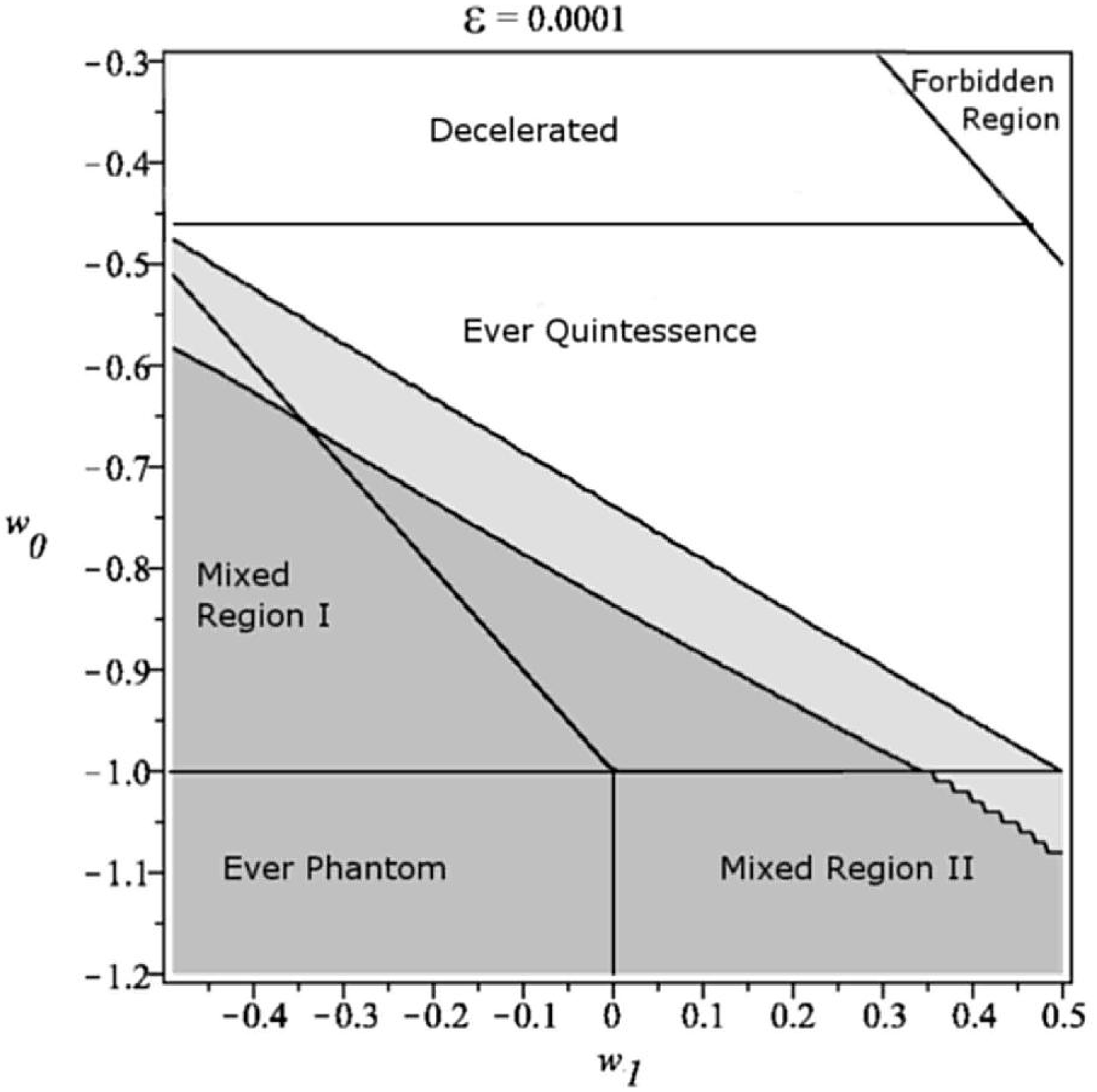}
\includegraphics*[scale=0.3,angle=0]{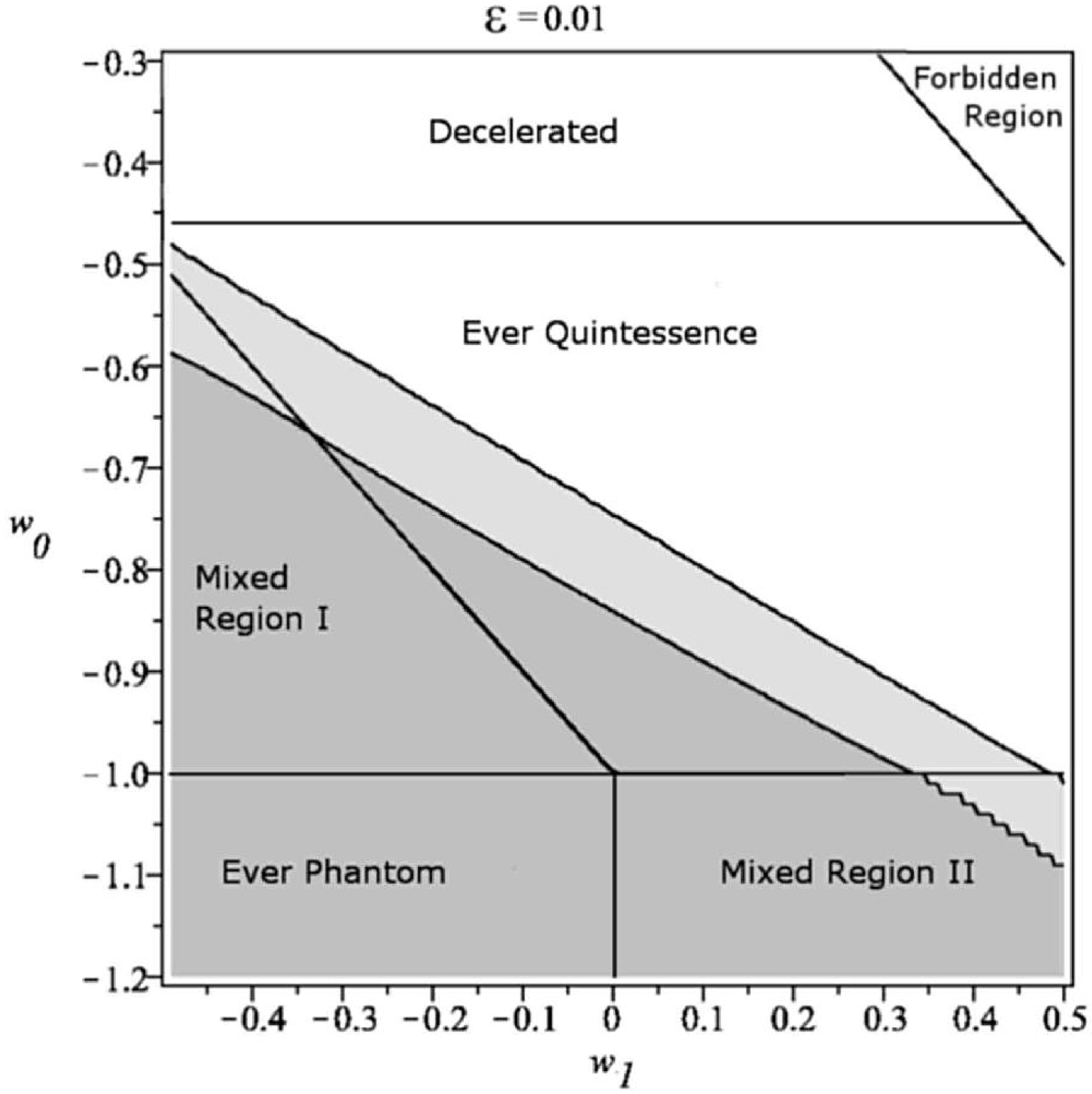}
\includegraphics*[scale=0.3,angle=0]{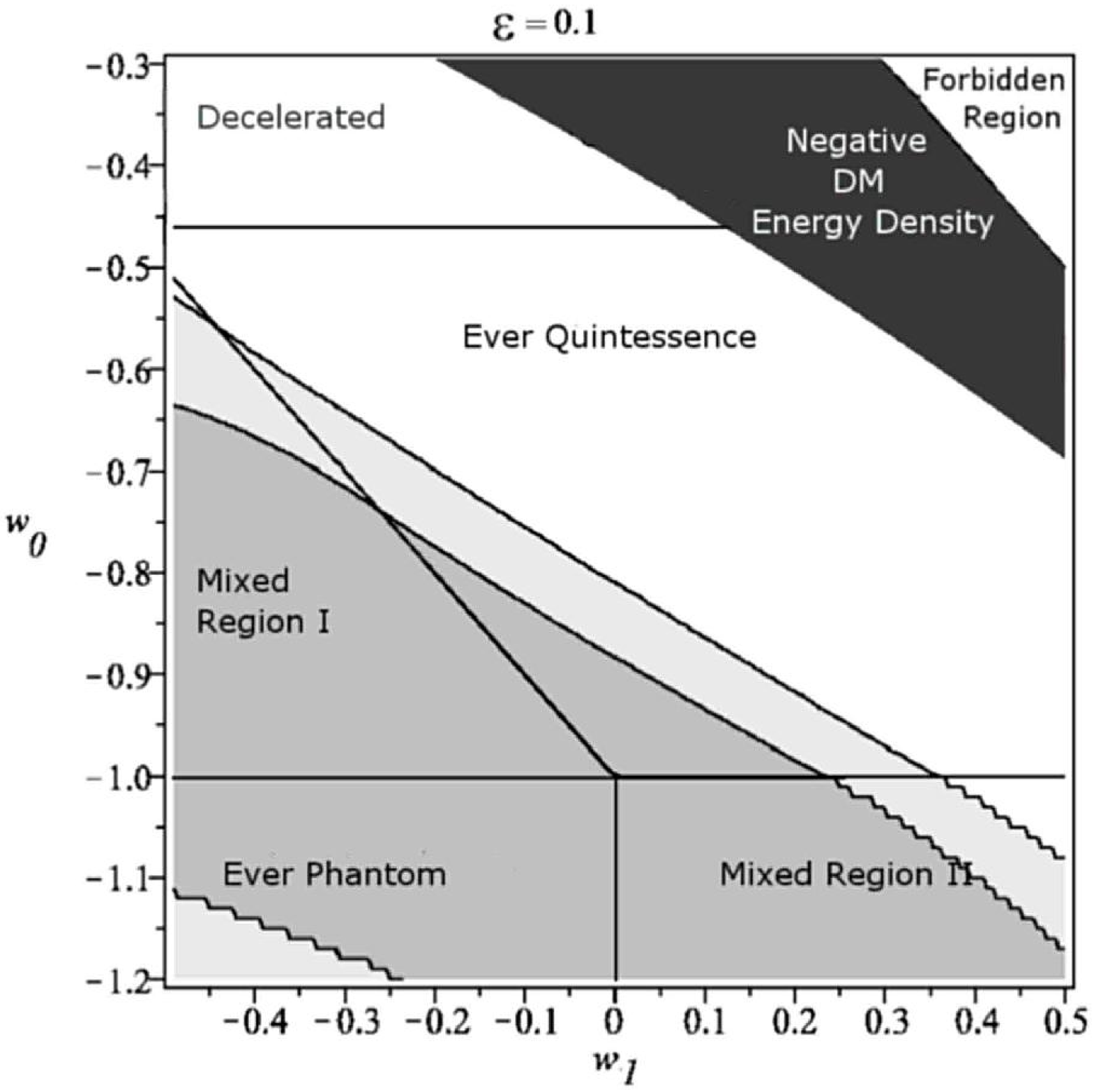}
\caption{In the light-gray sections of the four panels condition
(\ref{condition}) is met when use of both (\ref{int}.1) and the
Chevallier-Polarski-Linder parametrization (\ref{chevallier}) is
made. In the dark-gray sections, $|\Delta \phi|/ M_P< 0.7$; they
shrink with increasing $\epsilon$. The regions marked ``Ever
Quintessence" correspond to models such that $w_{0}+w_{1}> -1$ and
$w_{0}> -1$ for all $z >0$. Similarly, the regions marked ``Ever
Phantom" correspond to models such that $w(a) < -1$ for all $a <
1$. In the mixed region I  of each panel models transit from
phantom to quintessence as the Universe expands. In the mixed
region II they transit in the opposite sense. For $w_{0} > -0.46$
the expansion is decelerated at $a = 1$, as indicated in each
panel, see the text.} \label{fig3}
\end{figure}

Proceeding as before, we numerically evaluate $|\Delta \phi|/
M_{P}$ in terms of the parameters $\epsilon$, $w_{0}$ and $w_{1}$.
Figure \ref{fig3} shows $w_{0}$ vs. $w_{1}$ for different choices
of $\epsilon$. Points in the plane $(w_{1}, w_{0})$ satisfying
simultaneously $w_{0}+w_{1}> -1$ and $w_{0}> -1$ fulfill $w(a)
> -1$ for all $a >1 $ -i.e., they correspond to quintessence models
that never evolve into phantom. For small $\epsilon$ condition
(\ref{condition}) is violated  in a section of the quintessence
region, but only there. However, when $\epsilon$ takes moderate
values, the condition gets also violated by some models that
evolved from phantom to quintessence (some few models in the mixed
region I) and from some models that evolved in the opposite sense
(some models in mixed region II), see right bottom panel. Models
that always stay phantom  respect condition (\ref{condition}) for
all $\epsilon$  at any redshift $z \geq 0$. The triangular region
$w_{0}+w_{1} > 0$ at the top right corner of each panel is
observationally forbidden since points lying there correspond to
models that feature dark energy dominance at high redshifts. It is
noteworthy that it entirely falls in the region that violates
condition (\ref{condition}). Likewise when $\epsilon$ is about
$0.029$ or larger, an unphysical region of negative dark matter
density develops -again in the region that violates
(\ref{condition})-, see right bottom panel. It is also visible
that  $\Delta \phi$ slowly augments with $\epsilon$.

Also marked in each panel is the maximum $w_{0}$ for which there
is acceleration at $z = 0$. This value, $w_{0} = -0.46\,$, readily
follows by setting the present deceleration parameter, $q_{0} = -
\ddot{a}/(aH^{2})|_{z = 0} = \textstyle{1\over{2}}\, [1\, + \,
3w_{0}\, (1-\, \Omega_{b0}\,- \, \Omega_{c0})]$, to zero.

\begin{figure}[tbp]
\includegraphics*[scale=0.3,angle=0]{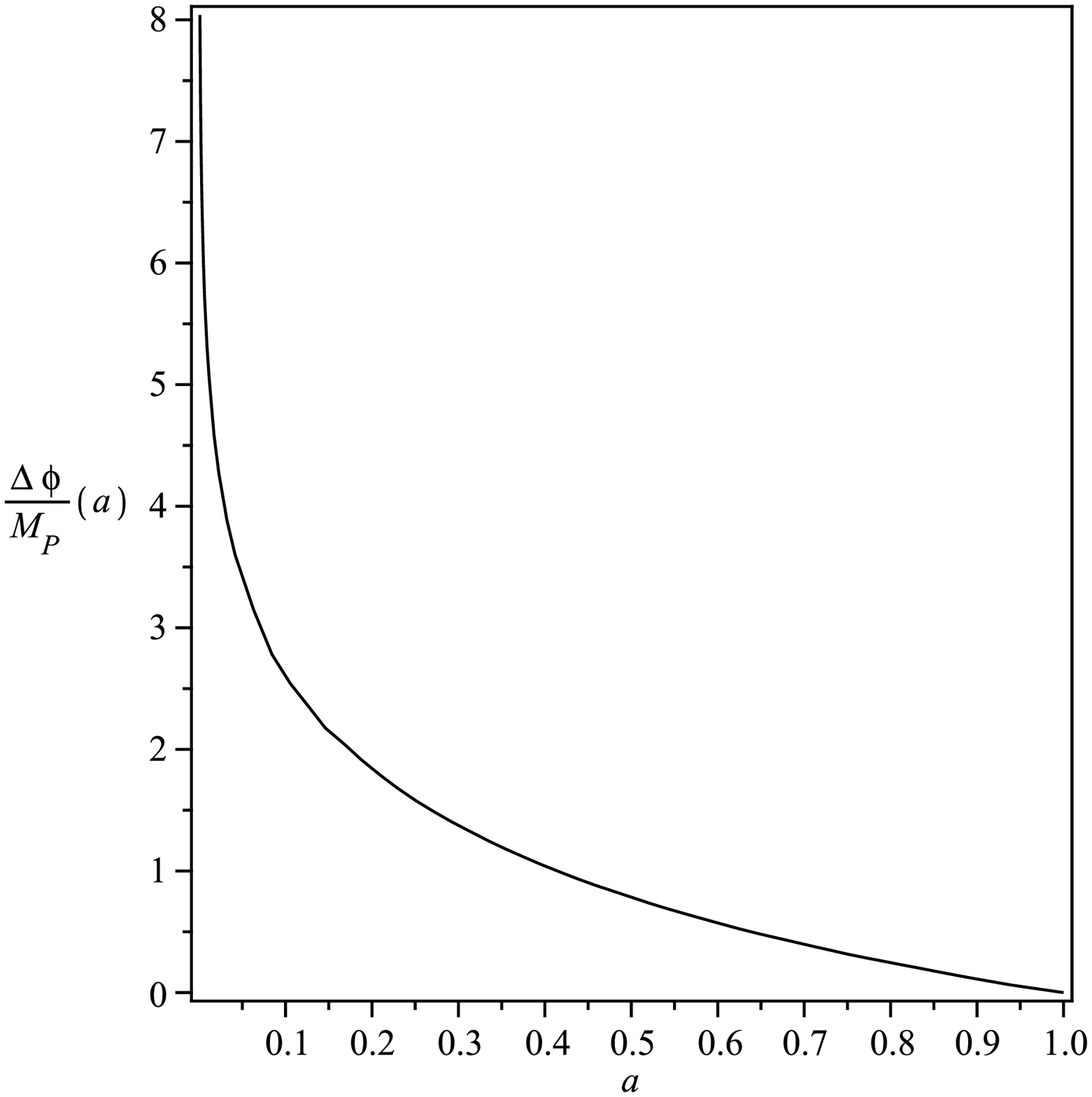}
\includegraphics*[scale=0.3,angle=0]{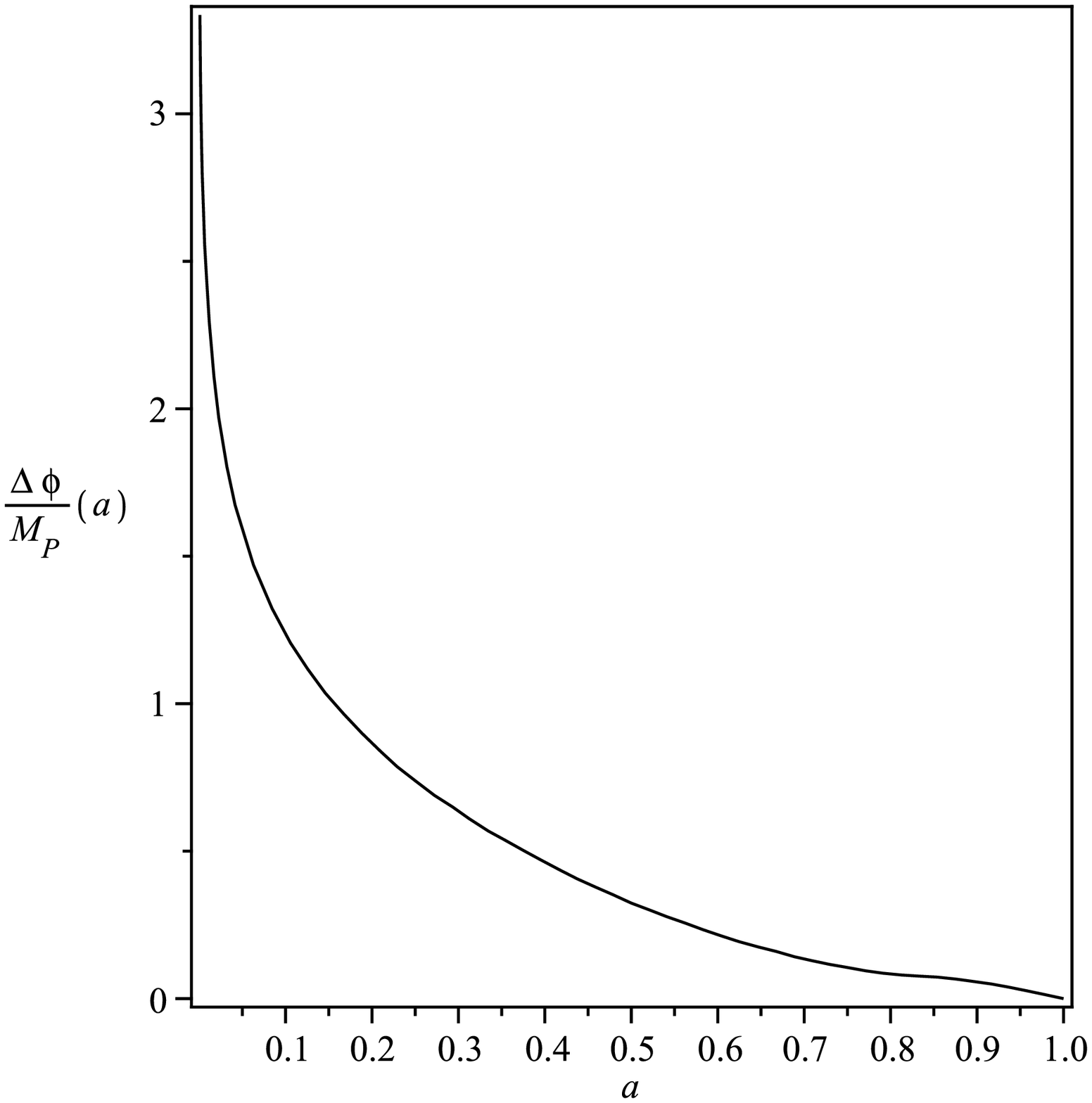}
\caption{$\Delta \phi/M_{P}$ in terms of the normalized scale
factor. In both panels $\epsilon = 0.01$ and the $w$ values
correspond to points on the straight border line of the forbidden
region of the left bottom panel of Fig. \ref{fig3}. Left panel:
$w_{0} = -0.46$. Right panel: $w_{0} = -1.2$.} \label{fig4}
\end{figure}

As we have checked, for $w$ values on the border of the forbidden
region the ratio $\Delta \phi/M_{P}$  is, at high redshifts (say,
$ z  >  20 $), consistently larger than 2 and can be as large as 7
-see Fig. \ref{fig4}. The fact that $w$ values in the forbidden
region correspond to dark energy dominance at early times and this
is incompatible with observation (e.g., the primeval
nucleosynthesis scenario and cosmic background radiation spectrum
would be very different from the one admitted today, and galaxies
could not have formed), suggests that, in general, scalar fields
respect the bound (\ref{condition}). This also holds true for the
forbidden regions of Figs. \ref{fig6}, \ref{fig9}, and \ref{fig11}
below.

\subsection{Barboza-Alcaniz parametrization}
As readily noted, Chevallier-Polarski-Linder's parametrization
(\ref{chevallier}) implies that $w(z)$ diverges as $z \rightarrow
-1$ (i.e., in the far future). To avoid this unpleasant feature
Barboza and Alcaniz proposed  the ansatz $w (z) = w_{0} \,+ \,
w_{1} \, \frac{z(1+z)}{1\, +\, z^{2}}$ or, equivalently, \be
w(a)=w_{0} \, +\, w_{1} \, \frac{1-a}{1-2a\, +\, 2a^{2}} \, ,
\label{alcaniz} \ee which ensures that $w(z)$ stays bounded in the
whole interval $-1 \leq z < \infty$ aside from behaving linearly
in redshift for $|z| \ll 1$ \cite{b-a}.

\begin{figure}[tbp]
\includegraphics*[scale=0.3,angle=0]{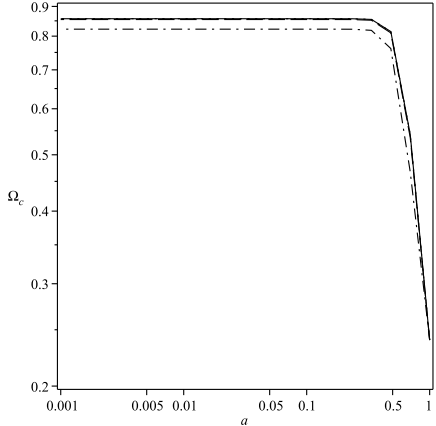}
\caption{Evolution of the fractional density of cold dark matter
$\Omega_{c} = \rho_{c}/(\rho_{b}+\rho_{c}+\rho_{\phi})$ in terms
of the normalized scale factor using the parametrization of $w(a)$
of Barboza and Alcaniz (\ref{alcaniz}) with $w_{0} = -1$ and
$w_{1} = -1.5$. Solid, dotted, dashed, and dot-dashed lines
correspond to $\epsilon=0,10^{-4}, 10^{-2},$ and $0.1$,
respectively; the three first practically overlap each other.}
\label{fig5}
\end{figure}

Assuming this novel parametrization alongside the interaction term
(\ref{int}.1), we get $ \rho_{\phi}= \rho_{\phi0}\,
a^{-3(1+w_0+w_1+\epsilon)}(1-2a+2a^2)^{3w_1/2}$. Again, $\rho_{c}$
must be calculated numerically. Figure \ref{fig5} illustrates the
behavior of $\Omega_{c}$.

\begin{figure}[tbp]
\includegraphics*[scale=0.3,angle=0]{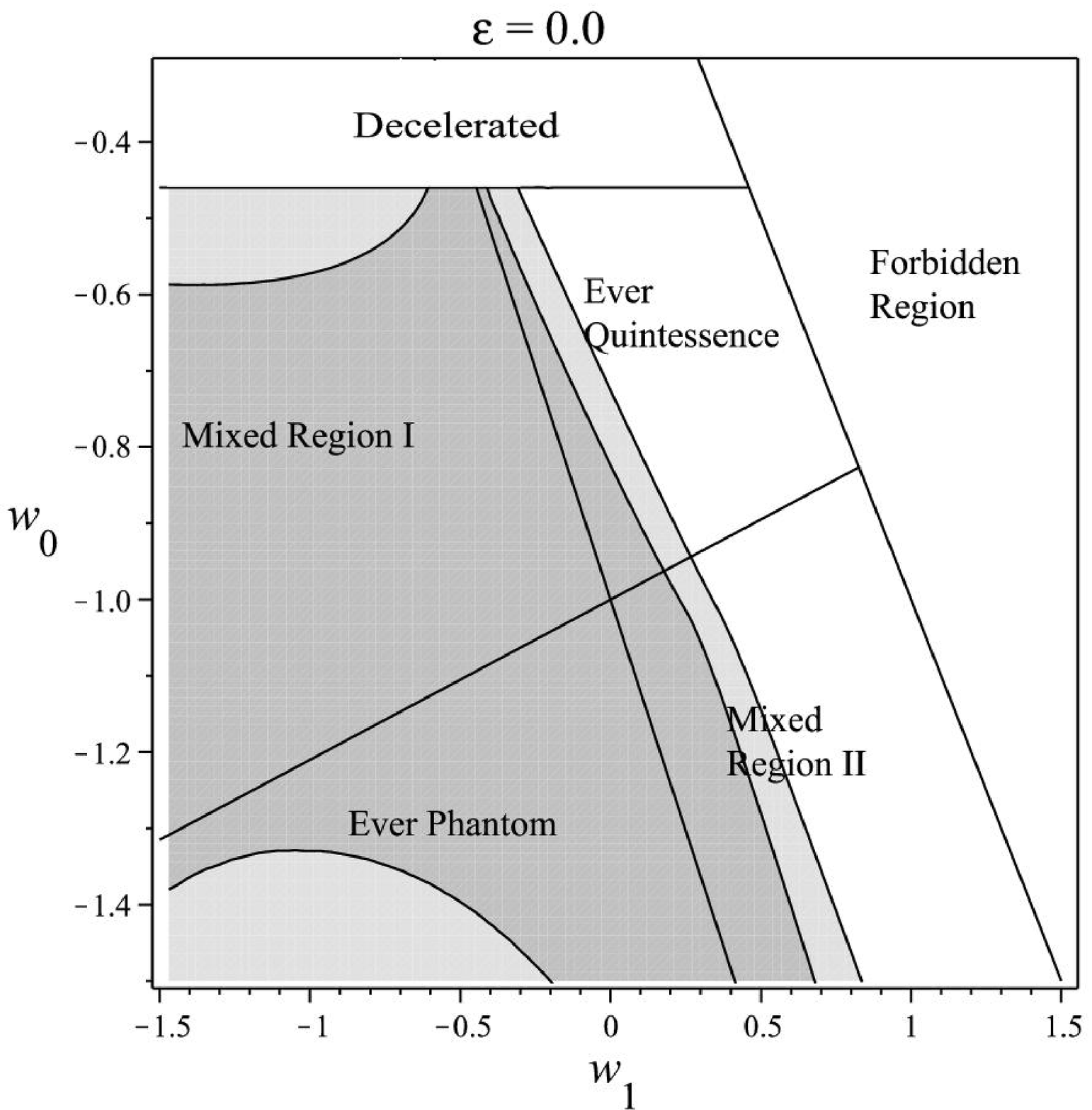}
\includegraphics*[scale=0.3,angle=0]{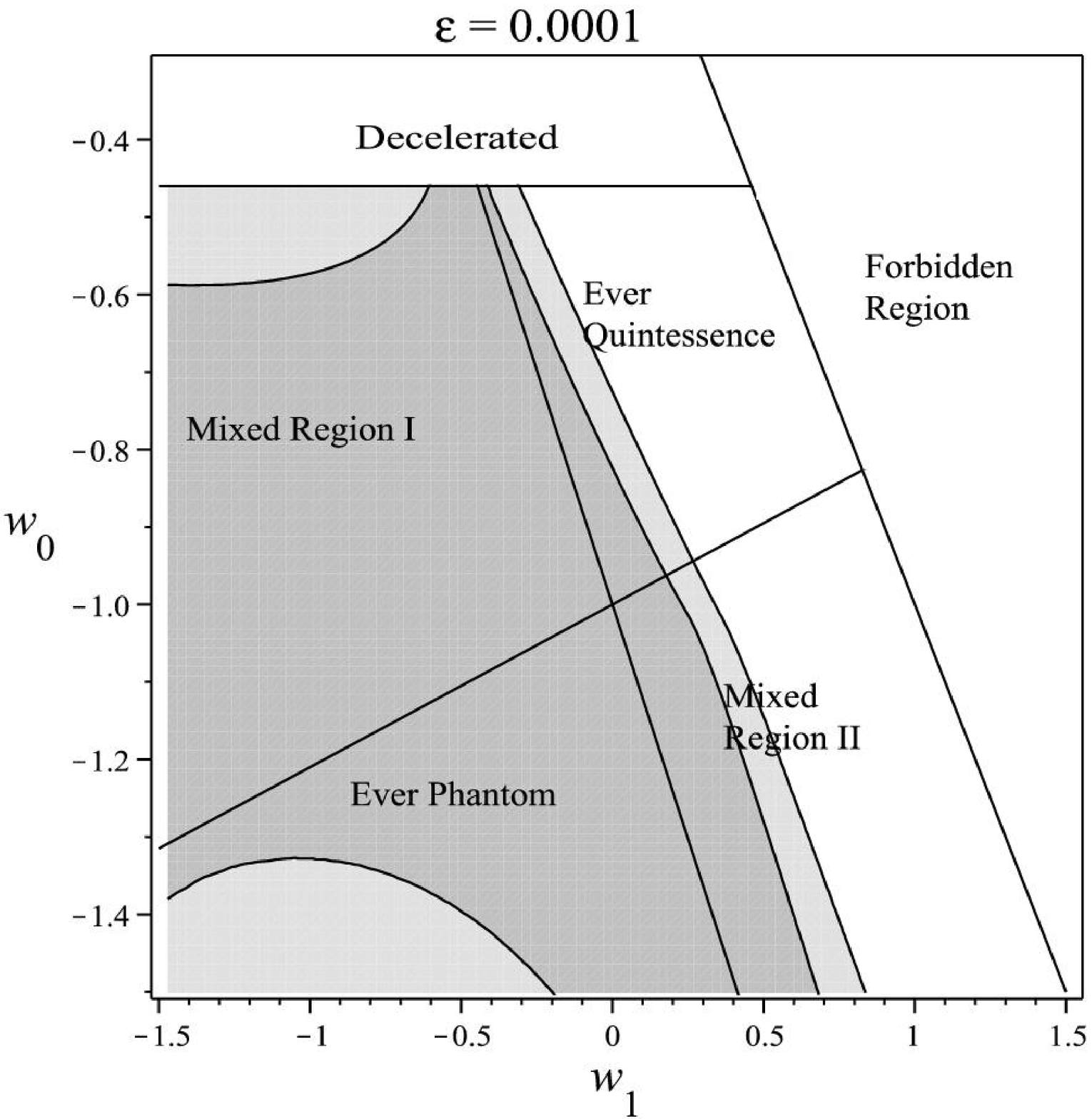}
\includegraphics*[scale=0.3,angle=0]{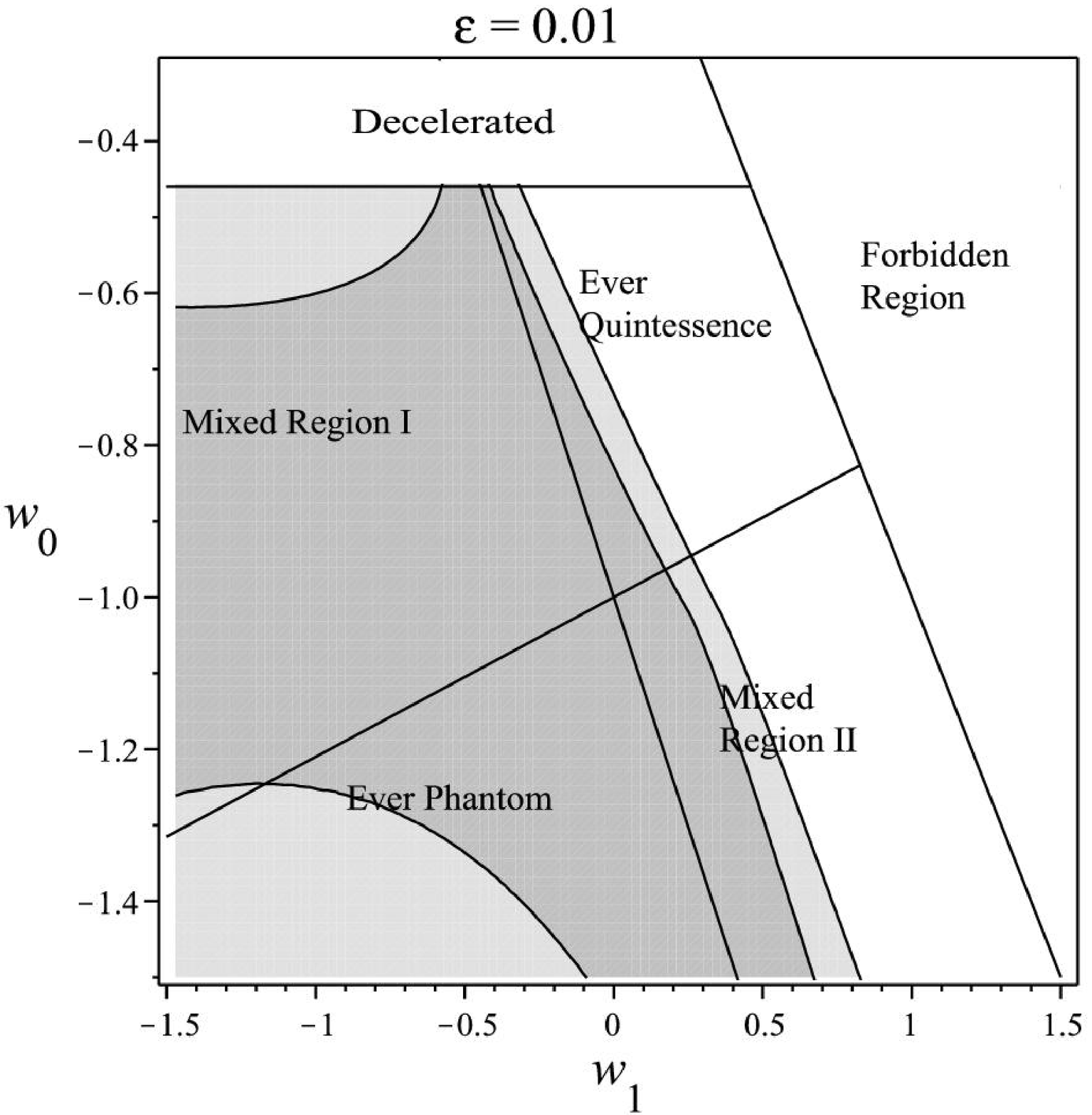}
\includegraphics*[scale=0.3,angle=0]{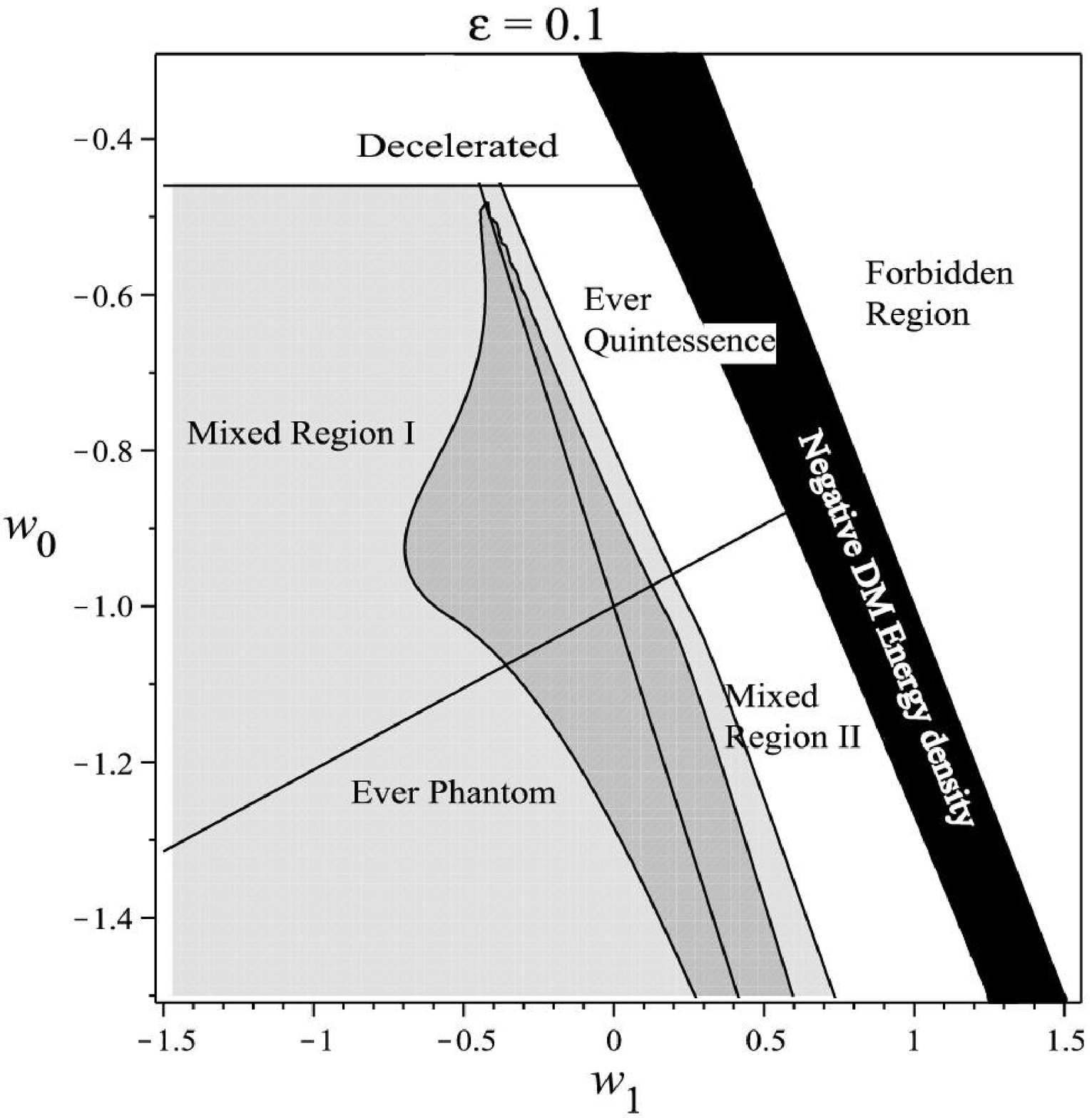}
\caption{Same as in Fig. \ref{fig3} except that $w(a)$ obeys the
parametrization of Barboza and Alcaniz, Eq. (\ref{alcaniz}).}
\label{fig6}
\end{figure}

Proceeding  as before, we numerically assess $|\Delta \phi|/ M_P$
in the redshift interval $(0, z_{ls})$. Figure \ref{fig6} depicts
$w_{0}$ vs. $w_{1}$ for various choices of $\epsilon$. If $w_{1}
> 0$, quintessence models lie in the region given by  $-1
\leq w_{0} - 0.21 \, w_{1}$ and $w_{0} + 1.21 \,  w_{1} \leq 1$;
if $w_{1} <0$, they lie in the region $-1 \leq w_{0} + 1.21 \,
w_{1}$ and $w_{0}- 0.21 \, w_{1} \leq 1$. Unlike the preceding
section, in the mixed regions II of the parameter space, phantom
models (that started as quintessence at high redshift) violate
condition (\ref{condition}) for whatever $\epsilon$. As before,
the triangular regions given by $w_{0} \, + \, w_{1} > 0$ are
observationally discarded as they correspond to dark energy
dominance at early times. These ones are much wider than in the
previous case. Again a region of negative cold dark matter density
develops, this time for $\epsilon$ values in excess of $\, 0.015\,
$ -see right bottom panel. As it is apparent, the interaction has
a noticeably effect on the variation of $\phi$ in the sense that
$\Delta \phi$ moderately augments with $\epsilon$. Comparison of
Figs. \ref{fig3} and \ref{fig6} reveals that the interaction
induces a bigger evolution of $\phi$ when $w(z)$ follows
Barboza-Alcaniz's parametrization than when it follows the
parametrization of Chevallier-Polarski-Linder.

\section{Interaction term proportional to the cold dark matter density}
In this section we take up $Q = 3 \, \epsilon H \rho_{c}$ for the
coupling  term and  resort to the expressions for the equation of
state parameter considered in the previous section.

\subsection{Constant $w$}
For $w = w_{0}$ we readily find
\begin{eqnarray}
\rho_{c} &=& \rho_{c0}\, a^{-3(1- \epsilon)} \, , \nonumber \\
\rho_{\phi} &=& \rho_{\phi0}\, a^{-3(1+w_{0})}\, + \, \rho_{c0} \,
\frac{\epsilon}{w_{0}+\epsilon} \, \left[ a^{-3(1+w_{0})}-
a^{-3(1-\epsilon)} \right].
\label{rho(a)2}
\end{eqnarray}

\begin{figure}[tbp]
\includegraphics*[scale=0.3,angle=0]{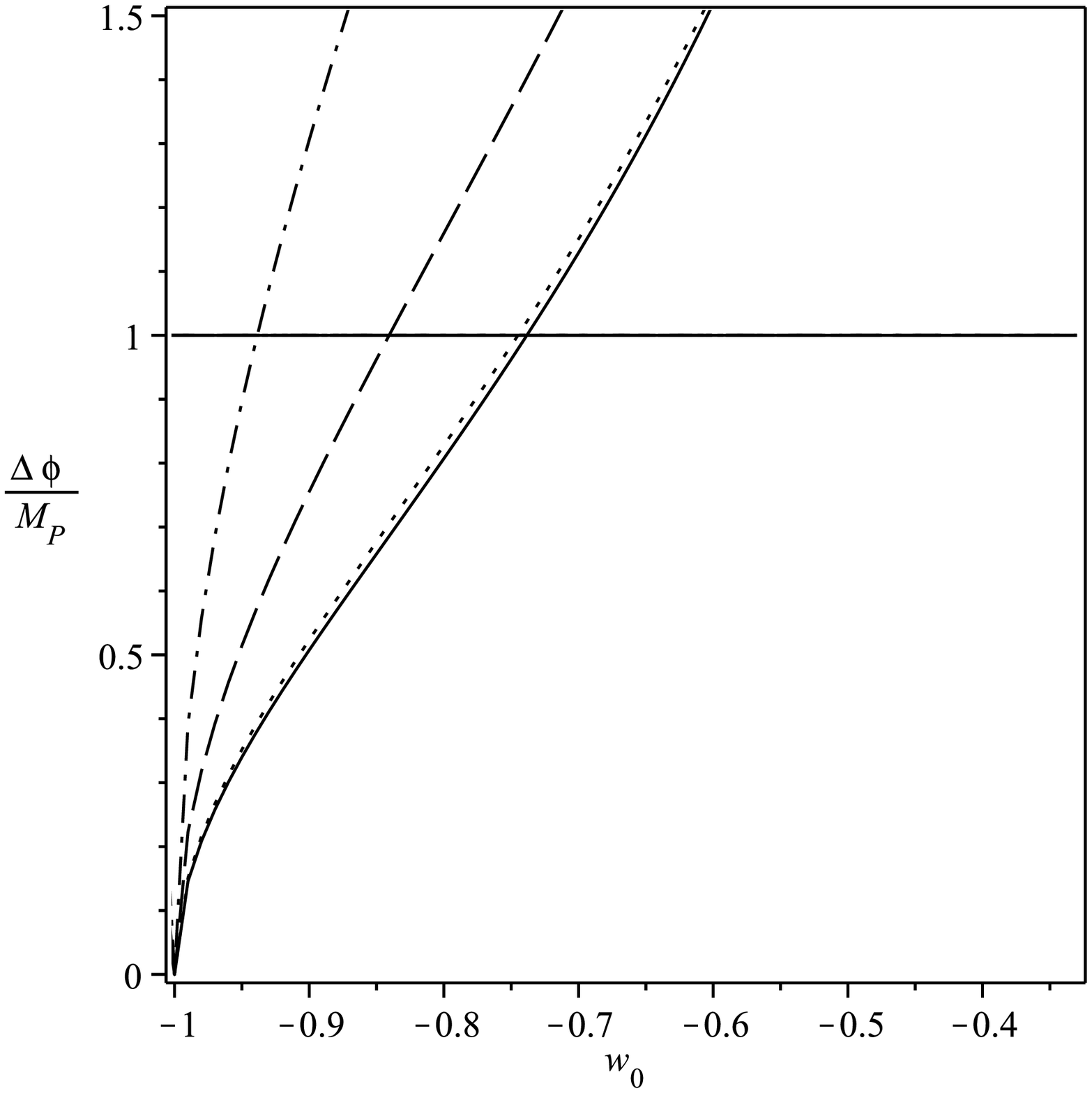}
\includegraphics*[scale=0.3,angle=0]{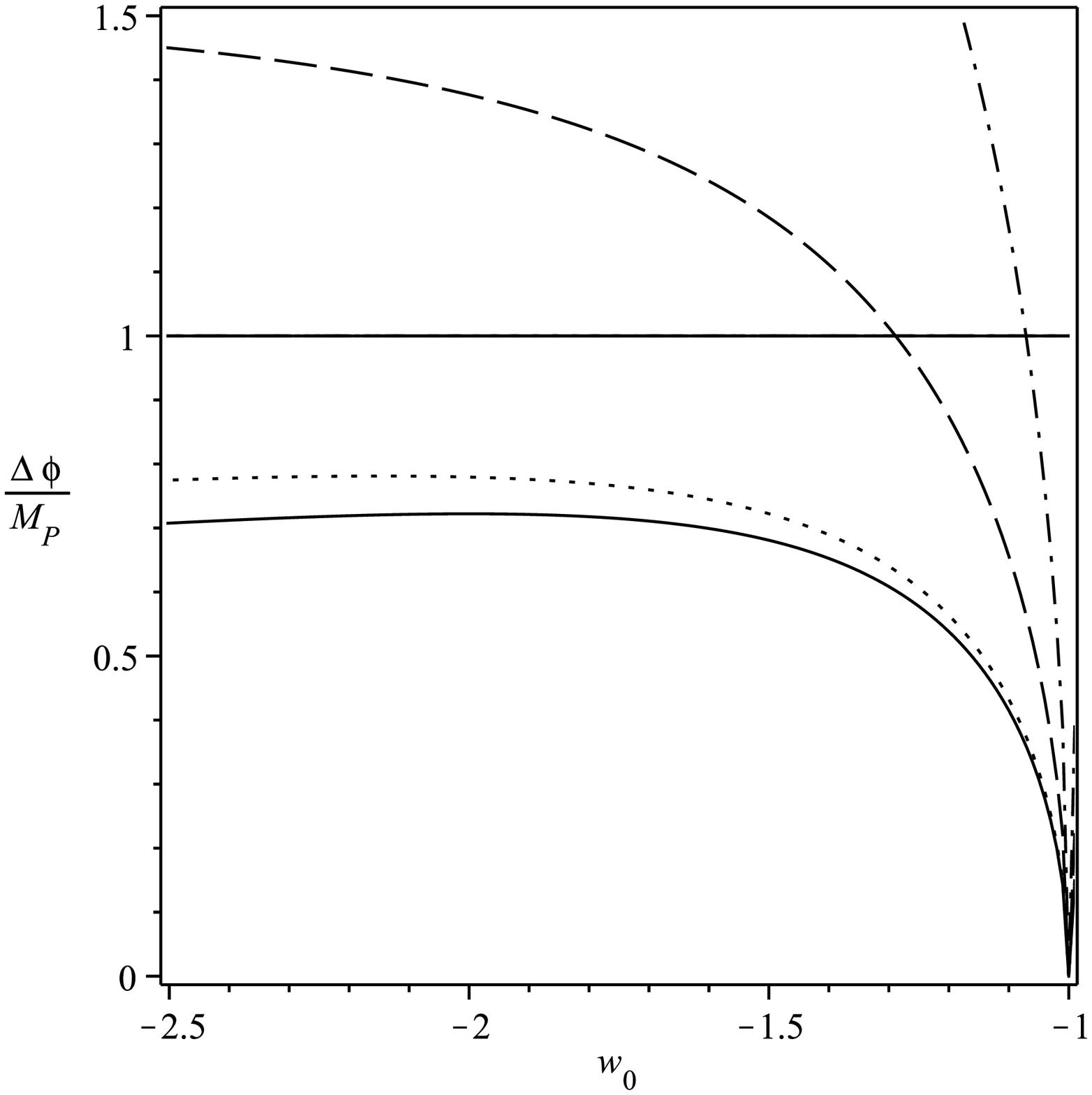}
\caption{$|\Delta \phi|/ M_P$ vs. $w_0$  for $Q$ given by
(\ref{int}.2) and constant $w$ evaluated in the redshift range
$(0, 1089)$. Left Panel: Quintessence solutions ($-1 <w_{0}
<-1/3$). Right panel: Phantom solutions ($w_0 < -1$). In both
panels solid, dotted, dashed, and dot-dashed lines correspond to
$\epsilon = 0, 10^{-4}, 10^{-2},$ and $0.1$, respectively.
Condition (\ref{condition}) implies $w_{0} < -0.738$ for $\epsilon
= 0$ and $w_{0} < -0.745$ for $\epsilon = 10^{-4}$; for $\epsilon
= 10^{-2}$ it implies $-1.299 < w_{0} < -0.841$, and $-1.071 <
w_{0} < -0.939$ for $\epsilon = 0.1$.} \label{fig7}
\end{figure}

Fig. \ref{fig7} shows $|\Delta \phi|/ M_P$ vs. $w_0$ for different
values of $\epsilon$. For quintessence  fields there is a range of
$w_{0}$ that violates condition (\ref{condition}) regardless the
value of $\epsilon$. For phantom fields the said condition only
holds when $\epsilon$ is very small. This contrast with
non-interacting phantom models which fulfill $|\Delta \phi|/ M_P <
1$ irrespective of $w_{0}$ (see Fig.1 of Ref. \cite{saridakis}).

\subsection{Chevallier-Polarski-Linder parametrization}
\begin{figure}[tbp]
\includegraphics*[scale=0.3,angle=0]{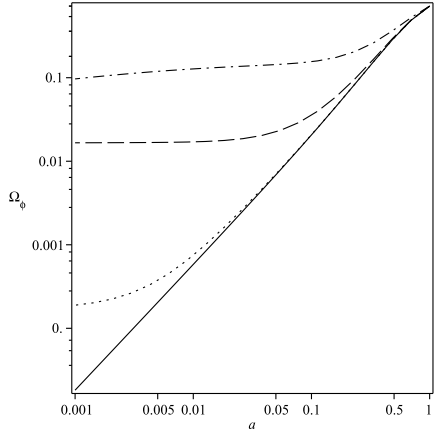}
\caption{Evolution of the fractional density of  dark energy for
the Chevallier-Polarski-Linder parametrization, Eq.
(\ref{chevallier}),   with $w_{0} = -1$ and $w_{1} = 0.5$. Solid,
dotted, dashed, and dot-dashed lines correspond to
$\epsilon=0,10^{-4}, 10^{-2},$ and $0.1$, respectively.}
\label{fig8}
\end{figure}

In this case $\rho_{c}=\rho_{c0}\, a^{-3(1-\epsilon)}$ while the
expression for $\rho_{\phi}$ must be found numerically. Figure
\ref{fig8} instances the evolution of $\Omega_{\phi}$ in terms of
the normalized scale factor.

\begin{figure}[tbp]
\includegraphics*[scale=0.3,angle=0]{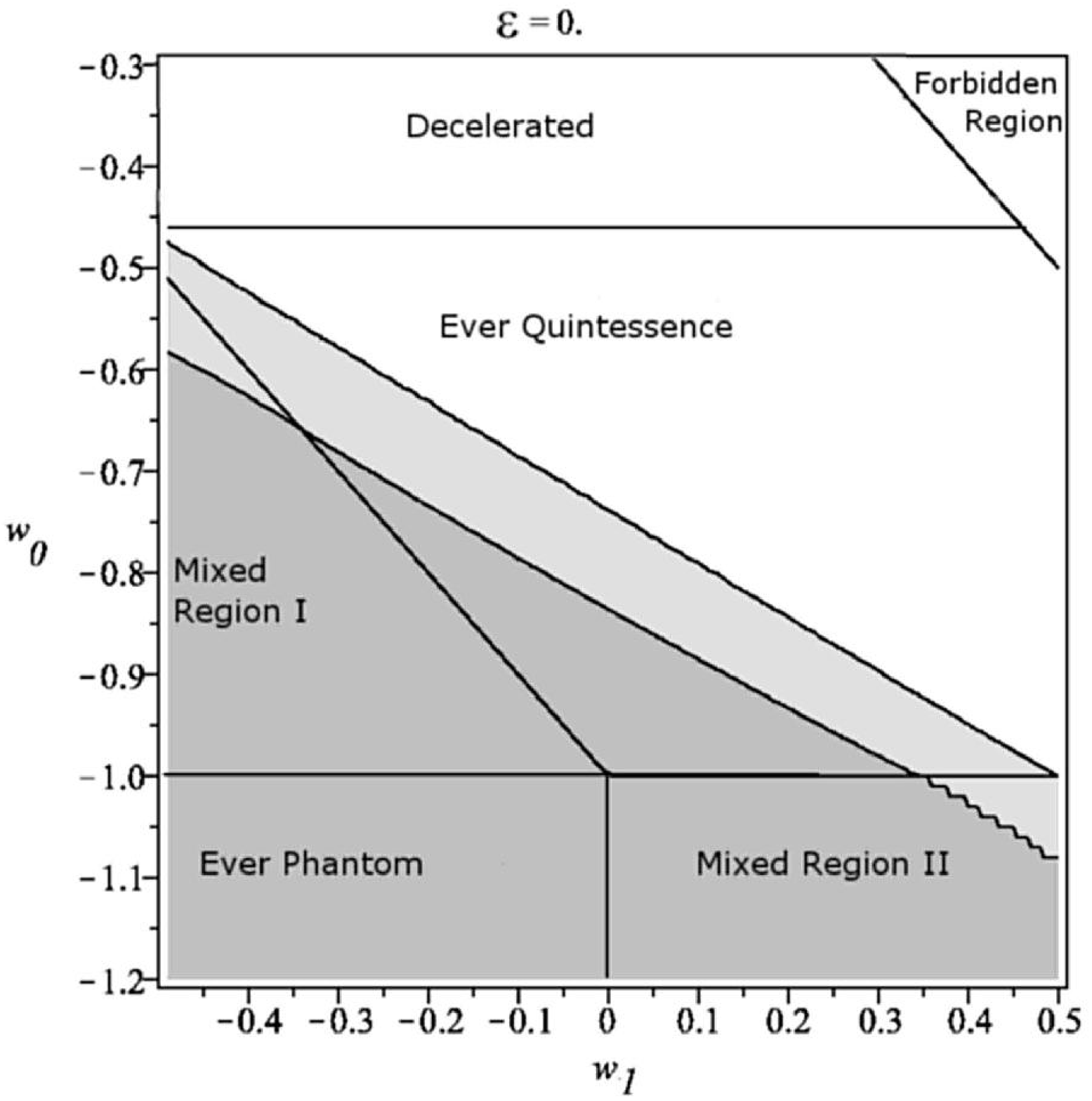}
\includegraphics*[scale=0.3,angle=0]{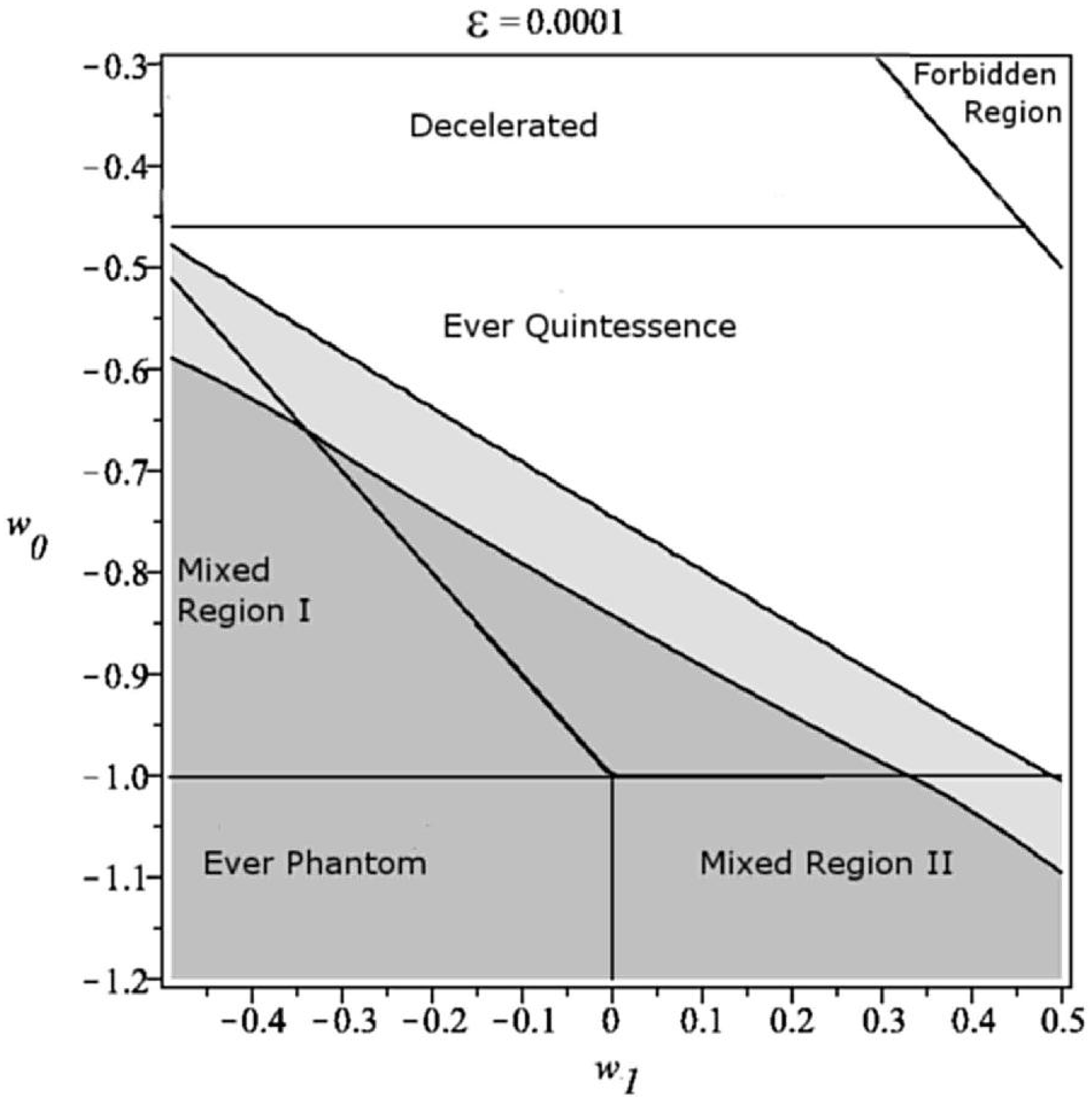}
\includegraphics*[scale=0.3,angle=0]{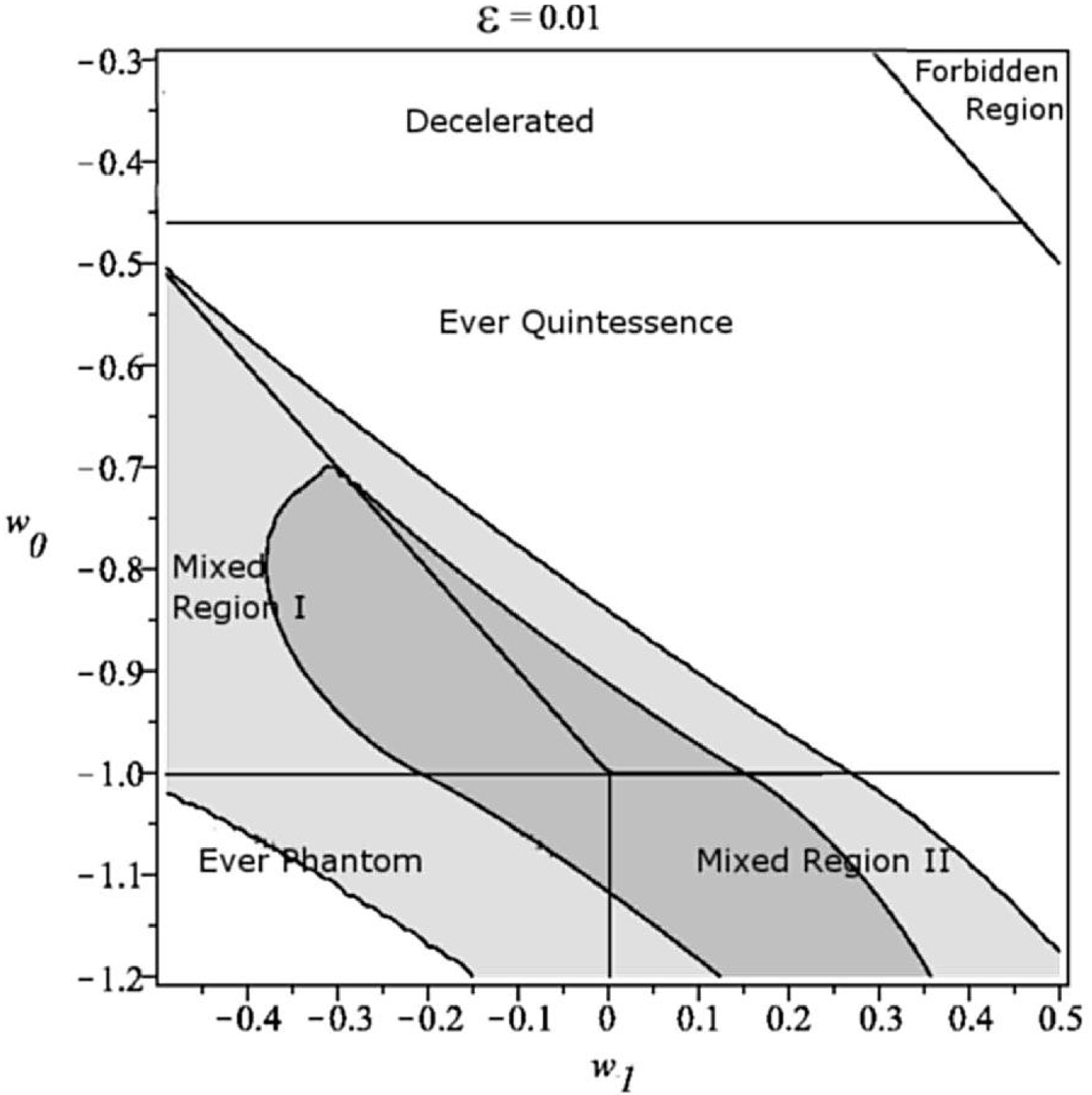}
\includegraphics*[scale=0.3,angle=0]{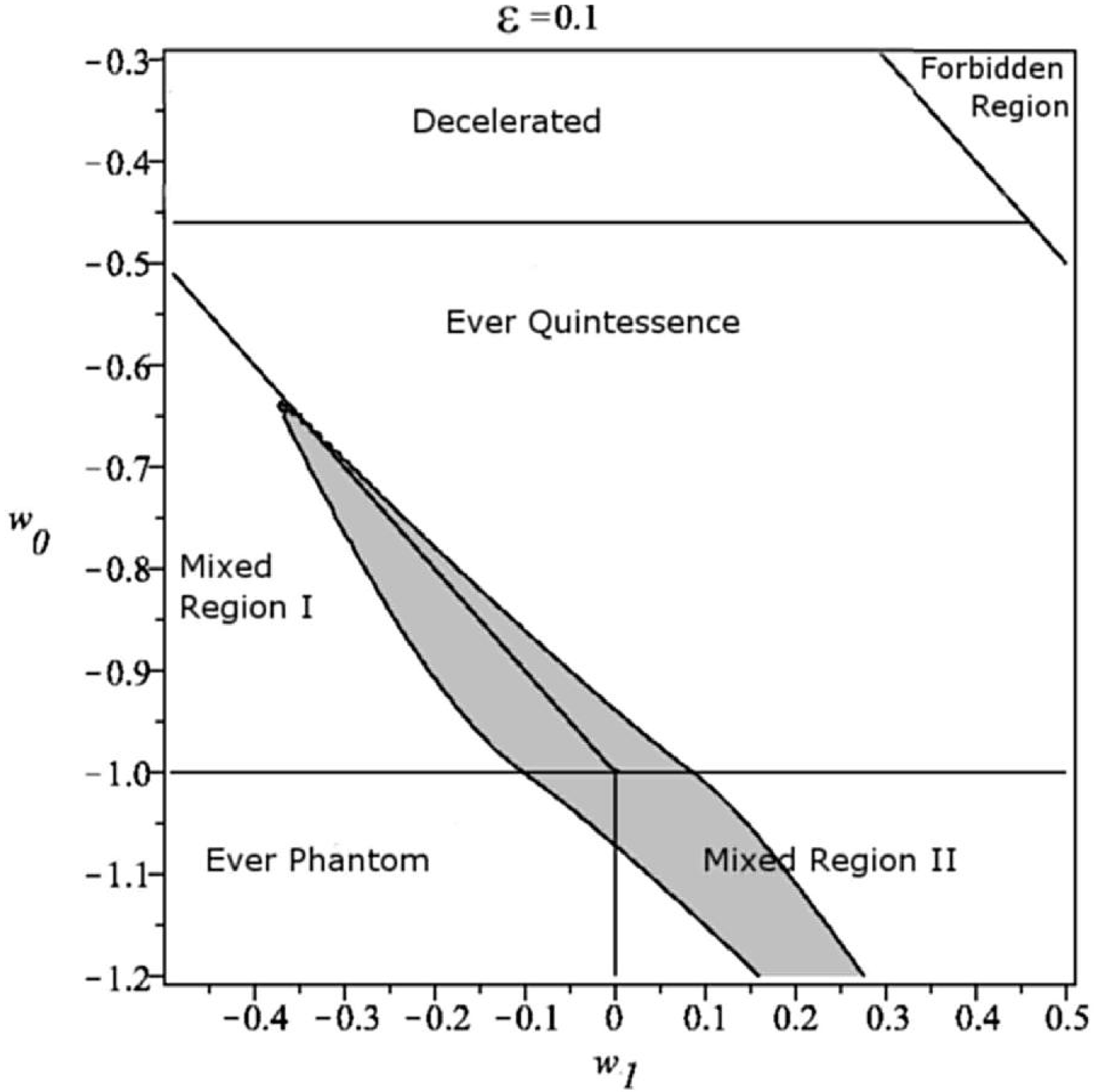}
\caption{Same as Fig. \ref{fig3} except that the interaction term
is now given by Eq. (\ref{int}.2) and that in the right bottom
panel only the $|\Delta \phi|/ M_P< 1$ region is depicted.}
\label{fig9}
\end{figure}

Dark energy models in Fig. \ref{fig9} such that $w_{0}+w_{1} > -1$
and $w_{0} > -1$ present a quintessence behavior during its whole
evolution. For small $\epsilon$ only models in this region can
violate condition (\ref{condition}). For larger $\epsilon$ values,
models in the other regions (bottom panels) can also violate it.
As it is manifest, the area of the offending region augments with
$\epsilon$. As in the preceding cases, models in the upper right
triangle of each panel are observationally discarded.

\subsection{Barboza-Alcaniz parametrization}
\begin{figure}[tbp]
\includegraphics*[scale=0.3,angle=0]{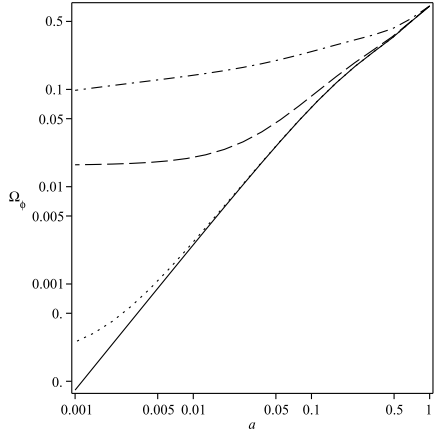}
\caption{Evolution of the fractional density of  dark energy  for
the Barboza-Alcaniz parametrization, Eq. (\ref{alcaniz}), with
$w_{0} = -1$ and $w_{1} = 0.5$. Solid, dotted, dashed, and
dot-dashed lines correspond to $\epsilon = 0, 10^{-4}, 10^{-2},$
and $0.1$, respectively.} \label{fig10}
\end{figure}

In this instance $\rho_{c}$ bears the same expression in terms of
the scale factor than in the previous case. Figure \ref{fig10}
illustrates the dependence of $\Omega_{\phi}$ on $a$.

\begin{figure}[tbp]
\includegraphics*[scale=0.3,angle=0]{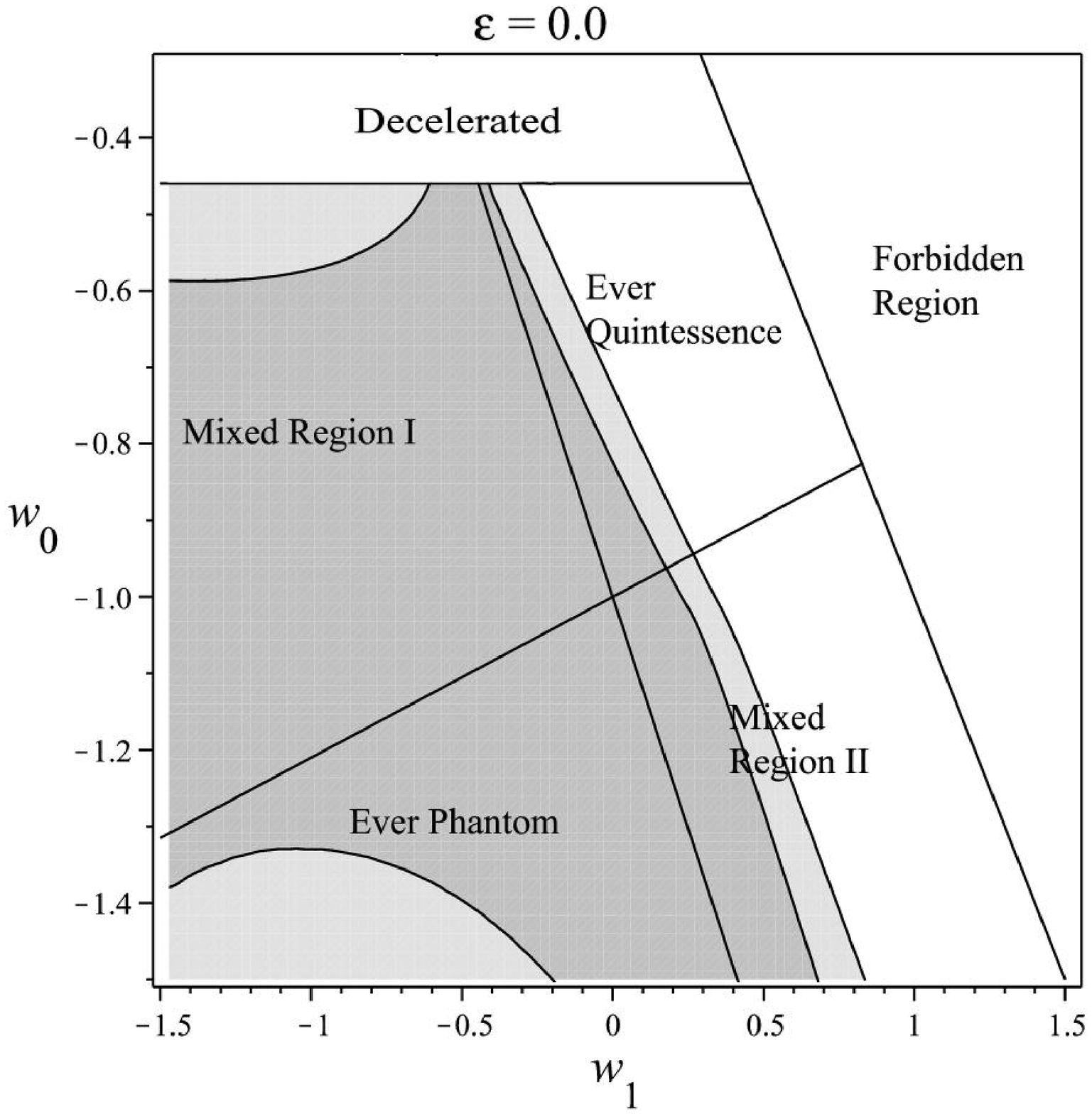}
\includegraphics*[scale=0.3,angle=0]{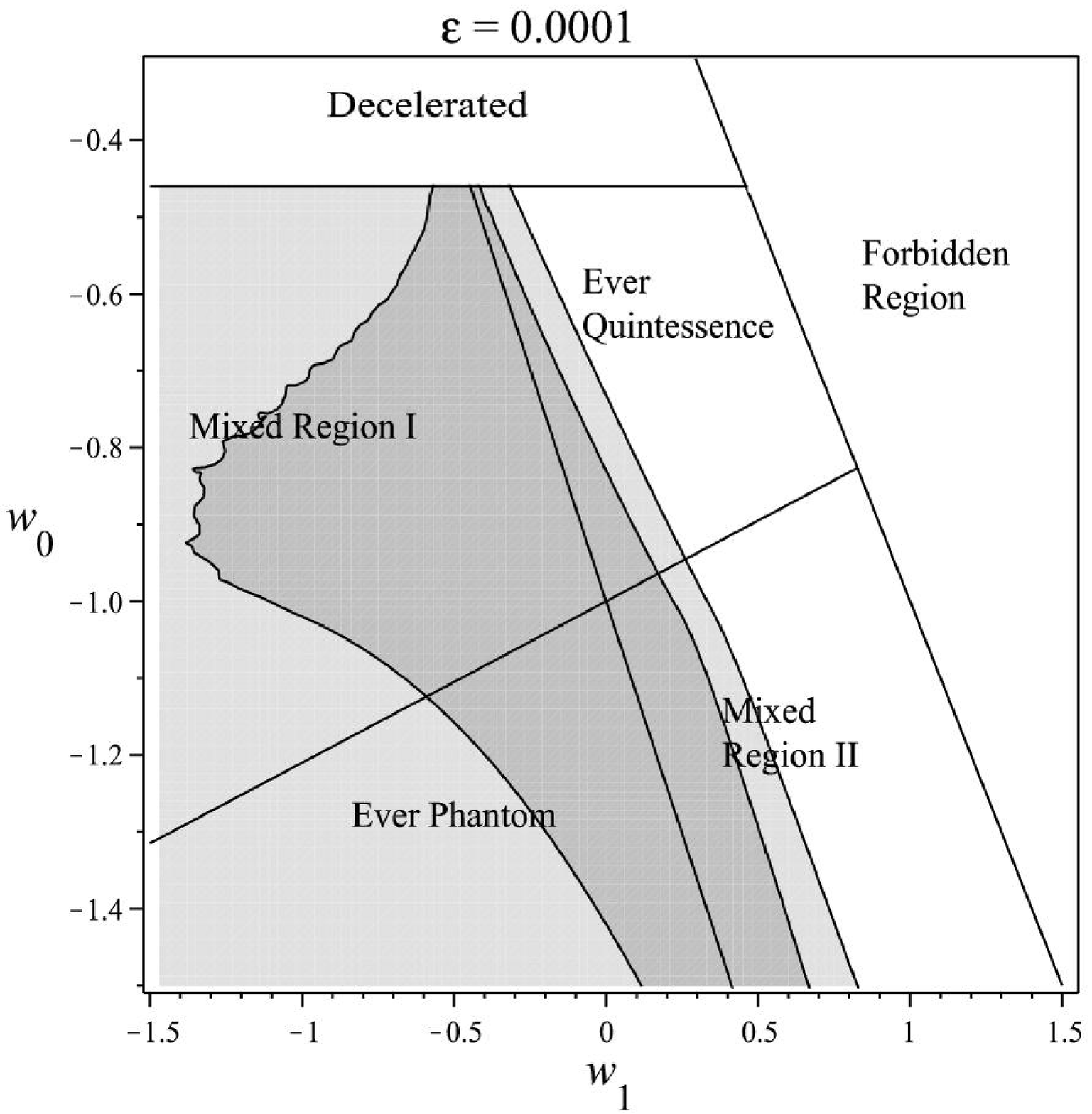}
\includegraphics*[scale=0.3,angle=0]{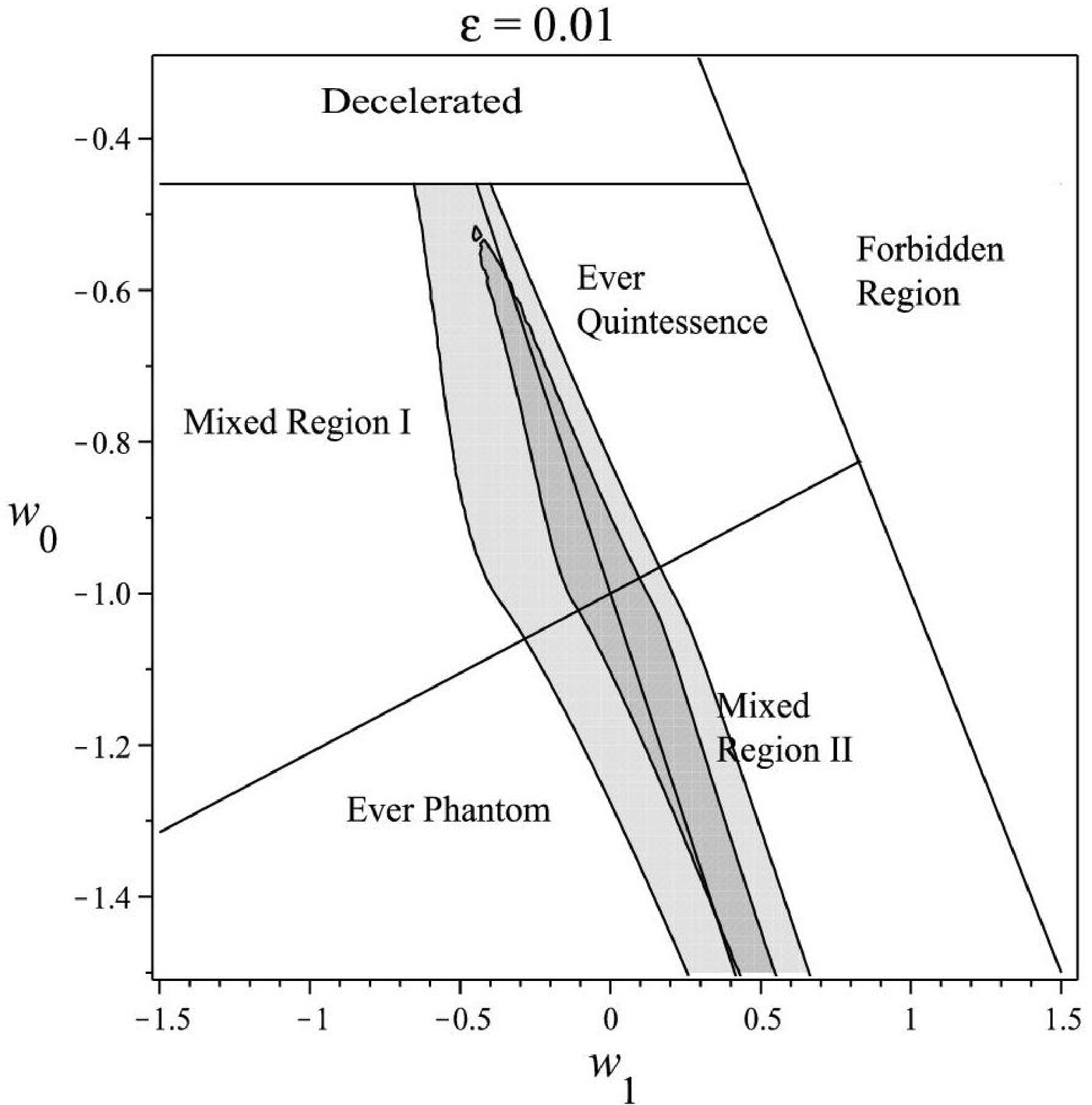}
\includegraphics*[scale=0.3,angle=0]{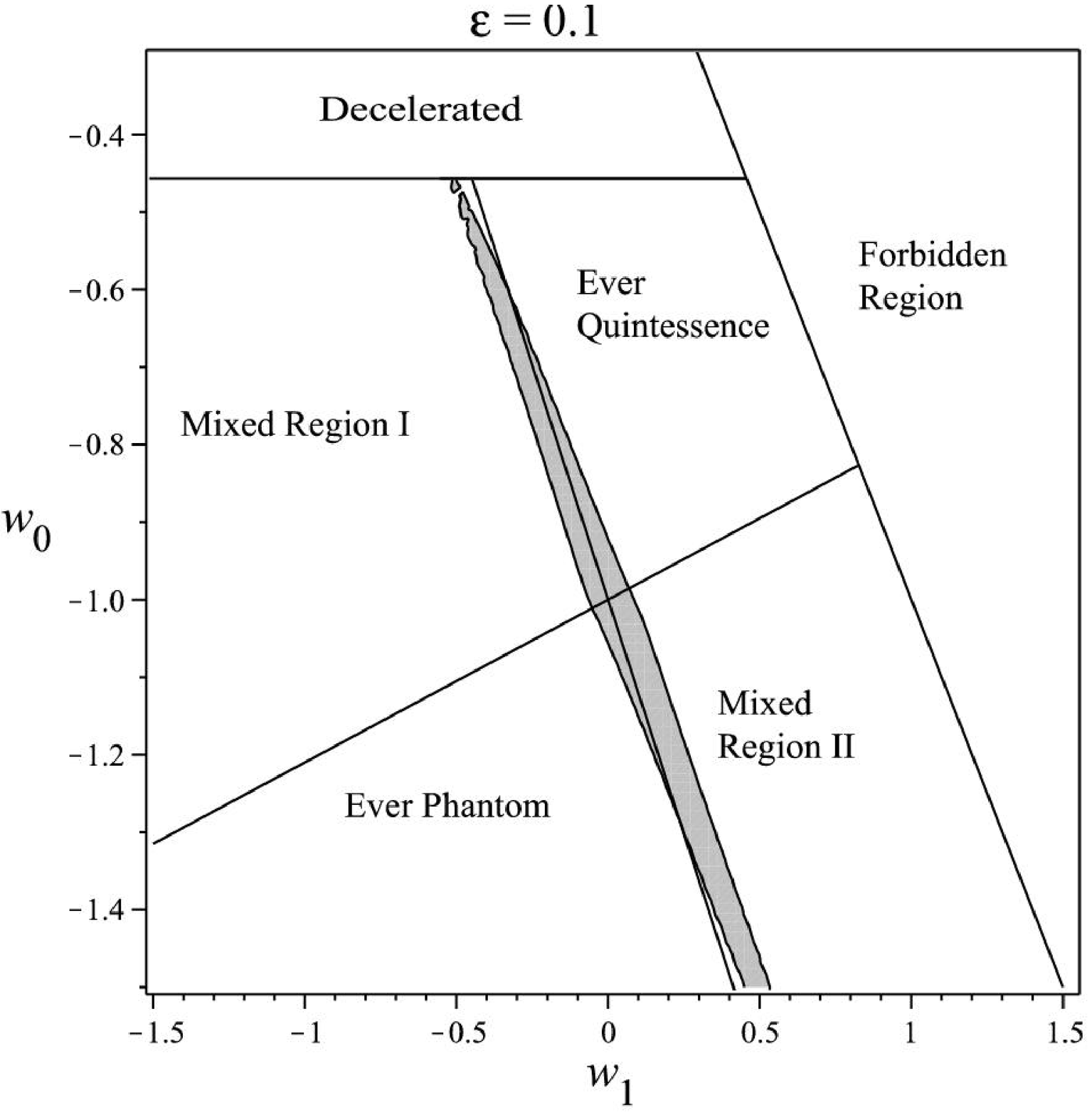}
\caption{Same as in Fig. \ref{fig9} except  that $w(a)$ obeys the
parametrization of Barboza and Alcaniz, Eq. (\ref{alcaniz}).}
\label{fig11}
\end{figure}

After having numerically evaluated $|\Delta \phi|/ M_{P}$, Fig.
\ref{fig11} displays $w_{0}$ vs. $w_{1}$ for different choices of
$\epsilon$ in the redshift interval $(0, z_{ls})$.
If $w_{1} > 0$, ever quintessence models belong to the region
given by $-1 \leq w_{0} - 0.21 \, w_{1}$ and $w_{0} + 1.21 \,
w_{1} \leq 1$; if $w_{1} < 0$, they lie in the region $-1 \leq
w_{0} + 1.21 \, w_{1}$ and $w_{0}- 0.21 \, w_{1} \leq 1$. Even for
small $\epsilon$, models in this region as well as models in the
mixed region II can breach condition (\ref{condition}) (top
panels). For larger $\epsilon$ values, models in the other regions
(bottom panels) can also breach it. As it is apparent from the
right bottom panel ($\epsilon = 0.1$), models that always stay
phantom are restricted to a very narrow region close to the line
$w_{0}=- 1 \, - \, 1.21 w_{1}$.

\section{Discussion}
Motivated by the reasonable assumption that the variation $\Delta
\phi$ of the field driving the present phase of cosmic accelerated
expansion should not exceed Planck's mass we have numerically
calculated the said evolution since last scattering till now for
different expressions of the equation of state parameter $w(a)$
and two couplings, $Q$, between dark matter and dark energy. This
constrains the parameter space $(w_{1}, w_{0})$ as shown in the
six cases studied. Figures \ref{fig1}, \ref{fig3}, \ref{fig6},
\ref{fig7}, \ref{fig9}, and \ref{fig11} explicitly exhibit the
regions of the parameter space $(w_{1}, w_{0})$ that fulfill the
bound $|\Delta \phi|/M_{P} < 1$. Our analysis extends those of
Huang \cite{huang} and Saridakis \cite{saridakis} who assumed that
dark matter and dark energy evolved separately and did not
consider Barboza and Alcaniz's parametrization \cite{b-a}.

The main results of this work can be summarized as follows:\\
\noindent $(i)$ When the interaction obeys (\ref{int}.1), models
that always stay phantom respect condition (\ref{condition}) at
any redshift and for any $\epsilon$ value. $(ii)$ Likewise,
depending on the values taken by $w_{0}$ and $w_{1}$ models that
always stay as quintessence as well as models that evolve from
quintessence to phantom and models that evolve in the opposite
sense can violate the said condition. $(iii)$ When $Q$ obeys
(\ref{int}.2) and $w(a)$ is given by the
Chevallier-Polarski-Linder parametrization (\ref{chevallier}), for
small $\epsilon$, all models that ever stay phantom and all models
in the mixed regions I and II respect bound (\ref{condition});
-top panels of Figs. \ref{fig9}. However, for not so small
$\epsilon$ values, models in these regions may also violate
(\ref{condition}) -bottom panels of Fig. \ref{fig9}. $(iv)$ When
$Q$ is given by (\ref{int}.2) and $w(a)$ obeys the Barboza-Alcaniz
parametrization (\ref{alcaniz}), for small $\epsilon$, all models
that ever stay phantom and models in the mixed region I satisfy
bound (\ref{condition}) -top panels of Fig. \ref{fig11}. By
contrast, for not so small $\epsilon$ values, models in these
regions may also violate (\ref{condition}) -bottom panels of Fig.
\ref{fig11}. $(v)$ The dark energy field $\phi$ appears to evolve
faster when: $(a)$ the coupling $Q$ depends explicitly on
$\rho_{c}$ than when it does on $\rho_{\phi}$, and $(b)$ when
$w(a)$ obeys the parametrization of Barboza-Alcaniz than when it
does that of Chevallier-Polarski-Linder. Point (a) is in line with
Caldera-Cabral {\it et al.}  result that interacting models with
$Q = 3 \, \epsilon \,  H \rho_{\phi}$ ``work better" than models
with $Q = 3 \, \epsilon \, H \rho_{c}$ \cite{gabriela1}. Further,
these models may help explain the non-vanishing temperature (about
$0.6$ Kelvin \cite{sterile}) of sterile neutrinos \cite{zhou}.

Our finding that $|\Delta \phi|$ increases with $\epsilon$
strengthens the view expressed in the Introduction  that $Q$ must
be small. Otherwise (except for (\ref{int}.1) and $w = {\rm
constant} < -1$, right panel of Fig. \ref{fig1}) condition
(\ref{condition}) would be violated for whatever $w$ not far from
$-1$.

We have confined ourselves to models in which the interaction term
is proportional to the Hubble factor (Eqs. (\ref{int})) whence our
results apply to them only. Other interaction terms can be found
in the literature, among others, $Q \propto \dot{\phi} \,
\rho_{c}$ \cite{among-others}, and $Q \propto (\rho_{c} \, + \,
\rho_{\phi})$ -see e.g. \cite{gabriela1, gabriela2}. However, the
former class of models present the drawback of being unable to
simultaneous lead to a correct sequence of cosmic eras (radiation,
matter, and dark energy) and solve the coincidence problem
\cite{challenges}, while the models we have considered do not
suffer from that \cite{drawback-free}. As for the latter class of
models, they generally lead to negative energy densities, either
of dark matter or dark energy, at early or late times
\cite{gabriela2}. Still, for some specific values of the
parameters entering the interaction these models are free of that
problem.

As is well known, due to quantum instabilities phantom fields may
find no place in Nature -see e.g. \cite{cline}. However, some
phantom models based on low-energy effective string theory may not
suffer from such pathologies \cite{federico} whence this issue is
not settled as yet. Moreover, higher derivative terms in the
Lagrangian may render phantom models stable \cite{creminelli}.

Before closing, it is sobering to recall that variable dark energy
fields are afflicted by the problem of their small mass ($m_{\phi}
\sim 10^{-33}$ eV). Due to this unpleasant feature they are looked
upon as not more natural than the cosmological constant, despite
the enormous fine tuning scale of the latter. Admittedly, this is
an unsolved problem that affects most (if not all) candidates of
evolving dark energy. However, recently, a supergravity based
approach that might point to the solution has been proposed
\cite{brax}. It provides a very small mass for the field and a
seemingly natural mechanism for a weak coupling between the field
and matter.

\acknowledgments{G.I. research was funded by the  ``Conseil
R\'{e}gional de la Bourgogne". This work has been partly supported
by the Spanish Ministry of Science and Innovation under Grant
FIS2009-13370-C02-01, and the ``Direcci\'{o} de Recerca de la
Gneralitat" under Grant 2009SGR-00164.}

\end{document}